\documentclass[aps,prd,superscriptaddress,floatfix,showpacs]{revtex4}

\usepackage{verbatim}

\usepackage{amsmath}
\usepackage{amsfonts}
\usepackage{amssymb}
\usepackage{graphicx}
\usepackage{bm}
\usepackage{color}
\usepackage{epsf}

\usepackage{psfrag}

\newcommand{\beq}[1]{
\begin{equation}\label{#1}}
\newcommand{\eeq}{\end{equation}}
\newcommand{\bea}[1]{
\marginpar{\small\textsf{#1}}
\begin{eqnarray}\label{#1}}
\newcommand{\eea}{\end{eqnarray}}
\def\bea{\begin{eqnarray}}
\def\eea{\end{eqnarray}}
\def\beas{\begin{eqnarray*}}
\def\eeas{\end{eqnarray*}}
\def\beqas{\begin{eqnarray*}}
\def\eqas{\end{eqnarray*}}
\def\beq{\begin{equation}}
\def\eeq{\end{equation}}
\def\beqd{\begin{displaymath}}
\def\eeqd{\end{displaymath}}
\def\eqd{\end{displaymath}}

\def\slashchar#1{\setbox0=\hbox{$#1$}
   \dimen0=\wd0
   \setbox1=\hbox{/} \dimen1=\wd1
   \ifdim\dimen0>\dimen1
      \rlap{\hbox to \dimen0{\hfil/\hfil}}
      #1
   \else\begin{eqnarray}
      \rlap{\hbox to \dimen1{\hfil$#1$\hfil}}
      /
   \fi}

\begin{document}
\title
{Exclusive neutrino-production of a charmed meson }
\author{ B.~Pire}
\affiliation{ Centre de Physique Th\'eorique, \'Ecole Polytechnique,
CNRS, Universit\'e Paris-Saclay, 91128 Palaiseau,     France }

\author{ L.~Szymanowski}
\affiliation{ National Centre for Nuclear Research (NCBJ), Warsaw, Poland}

\author{  J. Wagner}
\affiliation{ National Centre for Nuclear Research (NCBJ), Warsaw, Poland}
\date{\today}
\begin{abstract}

\noindent
We calculate the leading order in $\alpha_s$ QCD amplitude for exclusive  neutrino and antineutrino production of a $D$ pseudoscalar charmed meson on an unpolarized nucleon. We work in the framework of the collinear QCD approach where generalized parton distributions (GPDs) factorize from  perturbatively calculable coefficient functions. We include both $O(m_c)$ terms in the coefficient functions and $O(M_D)$ mass term contributions in the heavy meson distribution amplitudes. We emphasize the sensitivity of specific observables on the transversity quark GPDs.
\end{abstract}

\maketitle

\section{Introduction.}

The now well established framework of collinear QCD factorization \cite{fact1,fact2,fact3} for exclusive reactions mediated by a highly virtual photon in the generalized Bjorken regime describes hadronic amplitudes using generalized parton distributions (GPDs) which give access to a 3-dimensional analysis \cite{3d} of  the internal structure of hadrons.  
Neutrino production is another way to access (generalized) parton distributions  \cite{weakGPD,PS}. Although neutrino induced cross sections are orders of magnitudes smaller than those for electroproduction and neutrino beams are much more difficult to handle than charged lepton beams, they have  been very important to scrutinize the flavor content of the nucleon and the advent of new generations of neutrino experiments will open new possibilities. 
Using them would improve in a significant way future extraction of GPDs from the data \cite{extraction}. Moreover, charged current neutrino production is mediated by a massive vector boson exchange which is always highly virtual; one is thus tempted to apply a factorized description of the process amplitude down to small values of the momentum transfer $Q^2=-q^2$ carried by the $W^\pm$ boson.

Heavy quark production allows to extend the range of validity of collinear factorization, the heavy quark mass playing the role of the hard scale. Indeed kinematics (detailed below) shows that the relevant scale is $O(Q^2+m_c^2)$. Some data \cite{data} exist for charm production in medium and high energy neutrino and antineutrino  experiments and some specific channels ($D^0, D^\pm, D^*$) have been identified.

We shall thus write the  scattering amplitude $W ~N \to D ~N'$   in the collinear QCD framework as a convolution of  leading twist quark  and gluon GPDs with a  coefficient function calculated in the collinear kinematics taking heavy quark mass effects into account. This will allow a non-vanishing transverse amplitude $W_T ~N \to D ~N'$   with a leading contribution of order $\frac{m_c}{Q^2+m_c^2}$ \cite{PS}, built  from the convolution of chiral odd leading twist quark GPDs with a coefficient function of order $m_c$. In order to be consistent, we shall complement these leading terms with the order $\frac{M_D}{Q^2+M_D^2}$  contributions related to mass term in the distribution amplitudes of heavy mesons (see Eq. (\ref{DA})).

In this paper we consider the exclusive production of a pseudoscalar $D-$meson through the reactions on a proton (p) or a neutron (n) target:
\begin{eqnarray}
\nu_l (k)p(p_1) &\to& l^- (k')D^+ (p_D)p'(p_2) \,,\\
\nu_l (k)n(p_1) &\to& l^- (k')D^+ (p_D)n'(p_2) \,,\\
\nu_l (k)n(p_1) &\to& l^- (k')D^0 (p_D)p'(p_2) \,,\\
 \bar\nu_l (k) p(p_1) &\to& l^+ (k') D^-(p_D) p' (p_2)\,,\\
  \bar\nu_l (k) p(p_1) &\to& l^+ (k') \bar D^0(p_D) n' (p_2)\,,\\
  \bar\nu_l (k) n(p_1) &\to& l^+ (k') D^-(p_D) n' (p_2)\,,
\end{eqnarray}
in the kinematical domain where collinear factorization  leads to a description of the scattering amplitude 
in terms of nucleon GPDs and the $D-$meson distribution amplitude, with the hard subprocesses:
\begin{eqnarray}
W^+ d \to D^+ d~~~&,&~~~W^+ d \to D^0 u~~~~,~~~~ W^- \bar d \to  D^- \bar d~~~~,~~~~ W^- \bar d \to  \bar D^0 \bar u\,,
\end{eqnarray}
described by the  handbag Feynman diagrams of Fig.1 convoluted with chiral-even or chiral-odd quark GPDs, and the hard subprocesses:
\begin{eqnarray}
W^+ g \to D^+ g~~~~&,&~~~~ W^- g \to  D^- g\,,
\end{eqnarray}
convoluted with gluon GPDs (see Fig.2).

\begin{figure}
\includegraphics[width=0.8\textwidth]{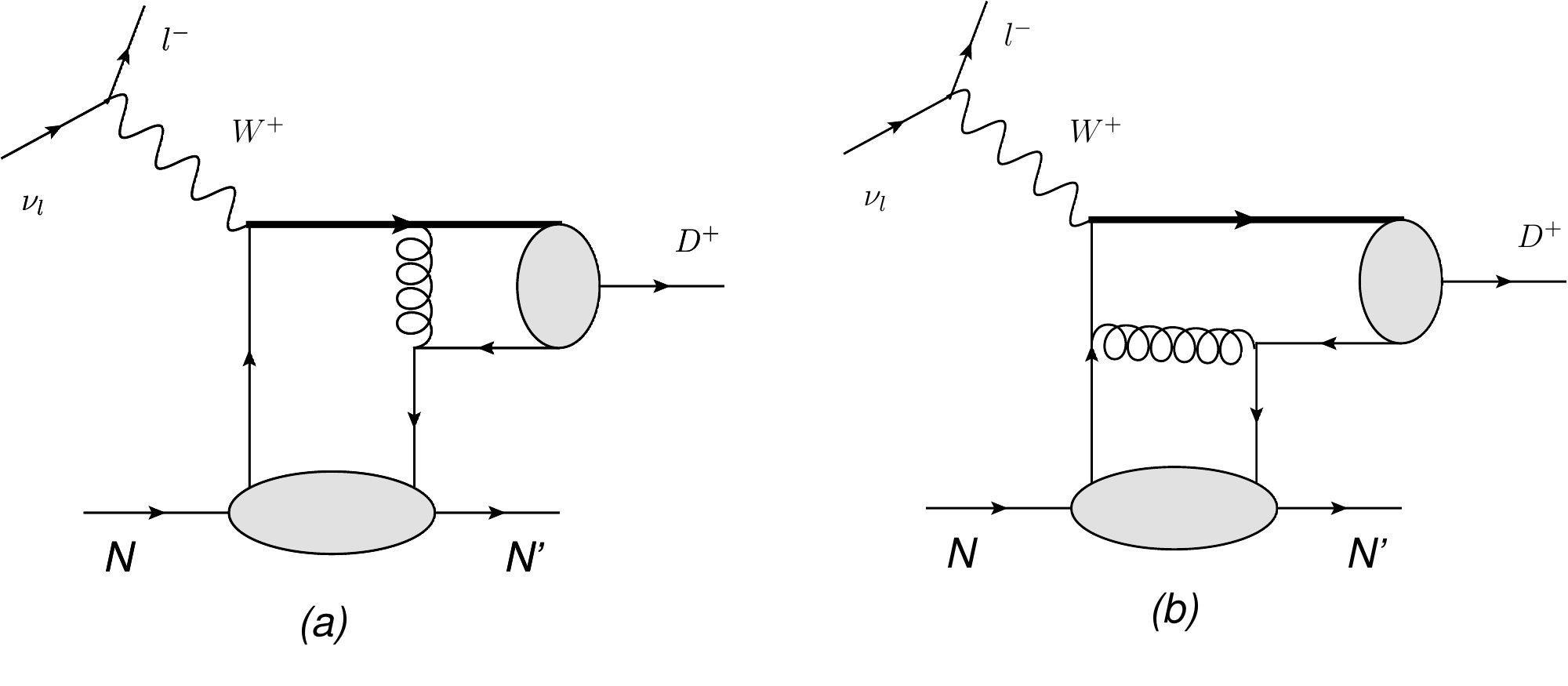}
\caption{Feynman diagrams for the factorized  amplitude for the $ \nu_l N \to l^-  D^+ N'$ or the $ \nu_l N \to l^-  D^0 N'$ process involving the quark GPDs; the thick line represents the heavy quark. In the Feynman gauge, diagram (a) involves convolution with both the transversity GPDs and the chiral even ones, whereas diagram (b) involves only chiral even GPDs.}
   \label{Fig1}
\end{figure}

\begin{figure}
\includegraphics[width=0.8\textwidth]{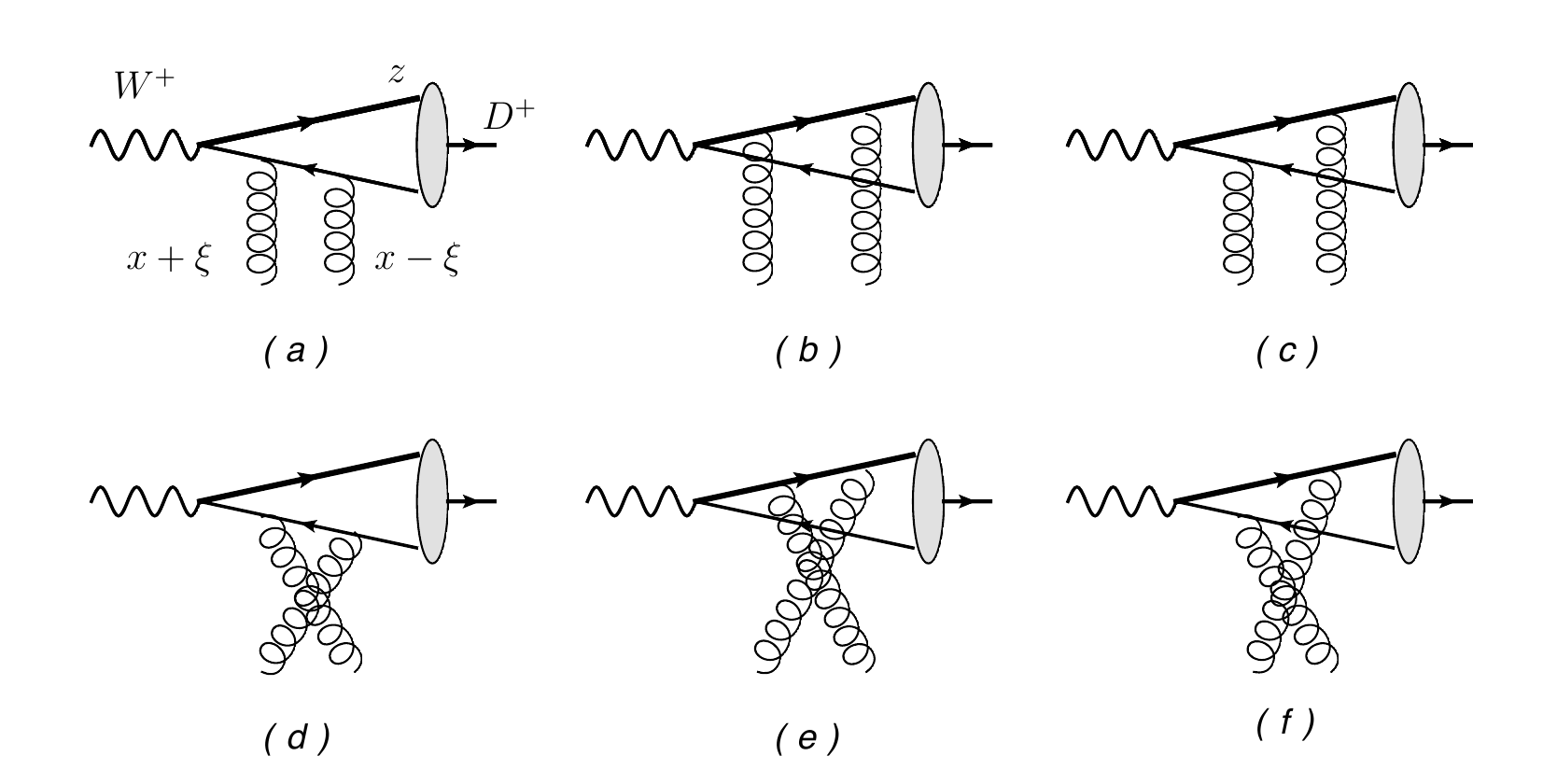}
\caption{Feynman diagrams for the factorized  amplitude for the $W^+ N \to D^+ N'$ process involving the gluon GPDs; the thick line represents the heavy quark.}
   \label{Fig2}
\end{figure}

Our kinematical notations are as follows ($m$ and $M_D$ are the nucleon and $D-$meson masses):
\begin{eqnarray}
&&q=k-k'~~~~~; ~~~~~Q^2 = -q^2~~~~~; ~~~~~\Delta = p_2-p_1 ~~~~~; ~~~~~\Delta^2=t \,;\nonumber\\
&&q^\mu= -2\xi' p^\mu +\frac{Q^2}{4\xi'} n^\mu ~;~\epsilon_L^\mu(q)= \frac{1}{Q} [2\xi' p^\mu +\frac{Q^2}{4\xi'} n^\mu] ~;~p_D^\mu=  2(\xi-\xi') p^\mu +\frac{M_D^2-\Delta_T^2}{4(\xi-\xi')}  n^\mu -\Delta_T^\mu \,; \\
&&p_1^\mu=(1+\xi)p^\mu +\frac{1}{2}  \frac{m^2-\Delta_T^2/4}{1+\xi} n^\mu -\frac{\Delta_T^\mu}{2}~~~~;~~~~ p_2^\mu=(1-\xi)p^\mu +\frac{1}{2}  \frac{m^2-\Delta_T^2/4}{1-\xi} n^\mu +\frac{\Delta_T^\mu}{2}
 \,,\nonumber
\end{eqnarray}
with $p^2 = n^2 = 0$ and $p.n=1$. 
As in  the double deeply virtual Compton scattering case~\cite{DDVCS}, it is meaningful to introduce two distinct momentum fractions:
 \begin{eqnarray}
\xi = - \frac{(p_2-p_1).n}{2} ~~~~~,~~~~~\xi' = - \frac{q.n}{2} \,.
\end{eqnarray} 
Momentum conservation leads to the relation:
\begin{equation}
\frac{Q^2}{\xi'}-\frac{\xi(4m^2-\Delta_T^2)}{1-\xi^2} = \frac{M_D^2-\Delta_T^2}{\xi-\xi'} \,.
\end{equation}
Neglecting the nucleon mass and $\Delta_T$, the approximate values of $\xi$ and $\xi'$ are
\begin{eqnarray}
\xi \approx \frac{Q^2+M_D^2}{4p_1.q-Q^2-M_D^2} ~~~~,~~~  \xi' \approx \frac{Q^2}{4p_1.q-Q^2-M_D^2} \,.
\end{eqnarray}
To unify the description of the scaling amplitude, we thus define a modified Bjorken variable
 \begin{eqnarray}
x_B^D \equiv \frac {Q^2+M_D^2}{2p_1.q} \neq x_B \equiv \frac {Q^2}{2p_1.q}\,,
\end{eqnarray}
which allows to express $\xi$ and $\xi'$ in a compact form:
 \begin{eqnarray}
\xi \approx \frac{x_B^D }{2-x_B^D } ~~~~~,~~~~~\xi' \approx \frac{x_B }{2-x_B^D } \,.
\end{eqnarray}
The difference  $\xi'-\xi = - \frac{p_D .n}{2}$  vanishes (in the strictly collinear case) when one neglects the meson mass. In this case, one gets the usual relations
\begin{eqnarray}
Q >>M_D ~~;~~\xi' \approx \xi ~~~~;~~~~ \xi \approx \frac{x_B}{2-x_B} \,.
\end{eqnarray}
On the other hand, if the meson mass is the relevant large scale (for instance in the limiting case where $Q^2$ vanishes as in the timelike Compton scattering kinematics \cite{TCS}) :
\begin{eqnarray}
Q^2\to0 ~~;~~\xi' \to 0 ~;~ \xi \approx \frac{\tau}{2-\tau} ~~;~~\tau = \frac{M_D^2}{s_{WN}-m^2}\,.
\end{eqnarray}

\section{Distribution amplitudes and GPDs} 
In the collinear factorization framework, the hadronization of the quark-antiquark pair is described by a distribution amplitude (DA) which obeys a twist expansion and evolution equations. Much work has been devoted to this subject \cite{heavyDA}.
The charmed meson distribution amplitudes are less known than the light meson ones. The limiting behavior of heavy mesons, which gives some constraints on the $B$ meson wave functions, is not very relevant for the charmed case. Here, we shall 
follow Ref. \cite{heavyDA2} and include some mass terms which will lead to order $\frac{M_D}{Q^2+M_D^2}$ contributions to the amplitudes;   omitting the path-ordered gauge link, the relevant distribution amplitude reads for the pseudo scalar $D^+$ meson :
\begin{eqnarray}
\langle D^+(P_D) | \bar c_\beta(y) d_\gamma(-y) |0 \rangle & =&
   i \frac{f_D}{4} \int_0^1 dz e^{i(z-\bar z)P_D.y} [(\hat P_D-M_D)\gamma^5]_{\gamma \beta}   \phi_D(z)\,,
   \label{DA}
         \end{eqnarray}
with $z=\frac{P_D^+-k^+}{P_D^+}$ and  where $\int_0^1 dz ~ \phi_D(z) = 1$,  $f_D= 0.223$ GeV. As usual, we denote $\bar z=1-z$ and  $\hat p = p_\mu \gamma^\mu$ for any vector $p$. 
         It has been argued that a heavy-light meson DA is strongly peaked around $z_0 = \frac {m_c}{M_D}$. We will parametrize  $\phi_D(z) $ as in Ref. \cite{heavyDA2}, {\it i.e.} $\phi_D(z) =6z(1-z)(1+C_D(2z-1))$ with $C_D\approx1.5$, which has a maximum around $z=0.7$.

We define the gluon and quark generalized parton distributions of a parton $q$ (here $q = u,\ d$) in the nucleon target   with the conventions of \cite{MD}.
To get the quantitative predictions for the neutrino-production observables, we use for the chirally even GPDs the Goloskokov-Kroll (G-K) model, based on the fits to deeply virtual meson production. Details of the model can be found in \cite{CEGPD}.

With respect to chiral odd twist 2 transversity GPDs, let us remind the reader that they correspond to the tensorial Dirac structure $\bar \psi^q\, \sigma^{\mu\nu} \,\psi^q  $. The leading GPD $H_T(x,\xi,t)$ is equal to the transversity PDF in the ($\xi=0~;~ t=0$) limit. The experimental access to these  GPDs  \cite{transGPDdef} has been  much discussed  \cite{transGPDno,transGPDacc} but much remains to be done. Models have been proposed \cite{models} and some  lattice calculation results exist \cite{lattice} for $H_T(x,\xi,t)$ and for 
the combination $\bar E_T(x, \xi, t) = 2\tilde H_T(x, \xi, t) +E_T(x, \xi, t)$. Since $\tilde E_T(x, \xi, t)$ is odd under $\xi \to -\xi$, most models find it vanishingly small. We will put it to zero in our numerical estimates of the observables. Since we are lacking decisive arguments about the relative sizes of $ \tilde H_T(x, \xi, t)$ and $E_T(x, \xi, t)$, we shall propose three quite extreme yet plausible models based on G-K parametrizations \cite{Goloskokov:2011rd} for $H_T(x, \xi, t) $ and  $ \bar E_T(x, \xi, t)$  :

\begin{itemize}
\item{model 1 : $\tilde H_T(x, \xi, t) = 0;  E_T(x, \xi, t) = \bar E_T(x, \xi, t) $.}
\item{model 2 : $\tilde H_T(x, \xi, t) = H_T(x, \xi, t) ;  E_T(x, \xi, t) = \bar E_T(x, \xi, t) -2 H_T(x, \xi, t) $.}
\item{model 3 : $\tilde H_T(x, \xi, t) = - H_T(x, \xi, t) ;  E_T(x, \xi, t) = \bar E_T(x, \xi, t) + 2 H_T(x, \xi, t) $.}
\end{itemize}
\begin{figure}
\centering
\includegraphics[width=15cm]{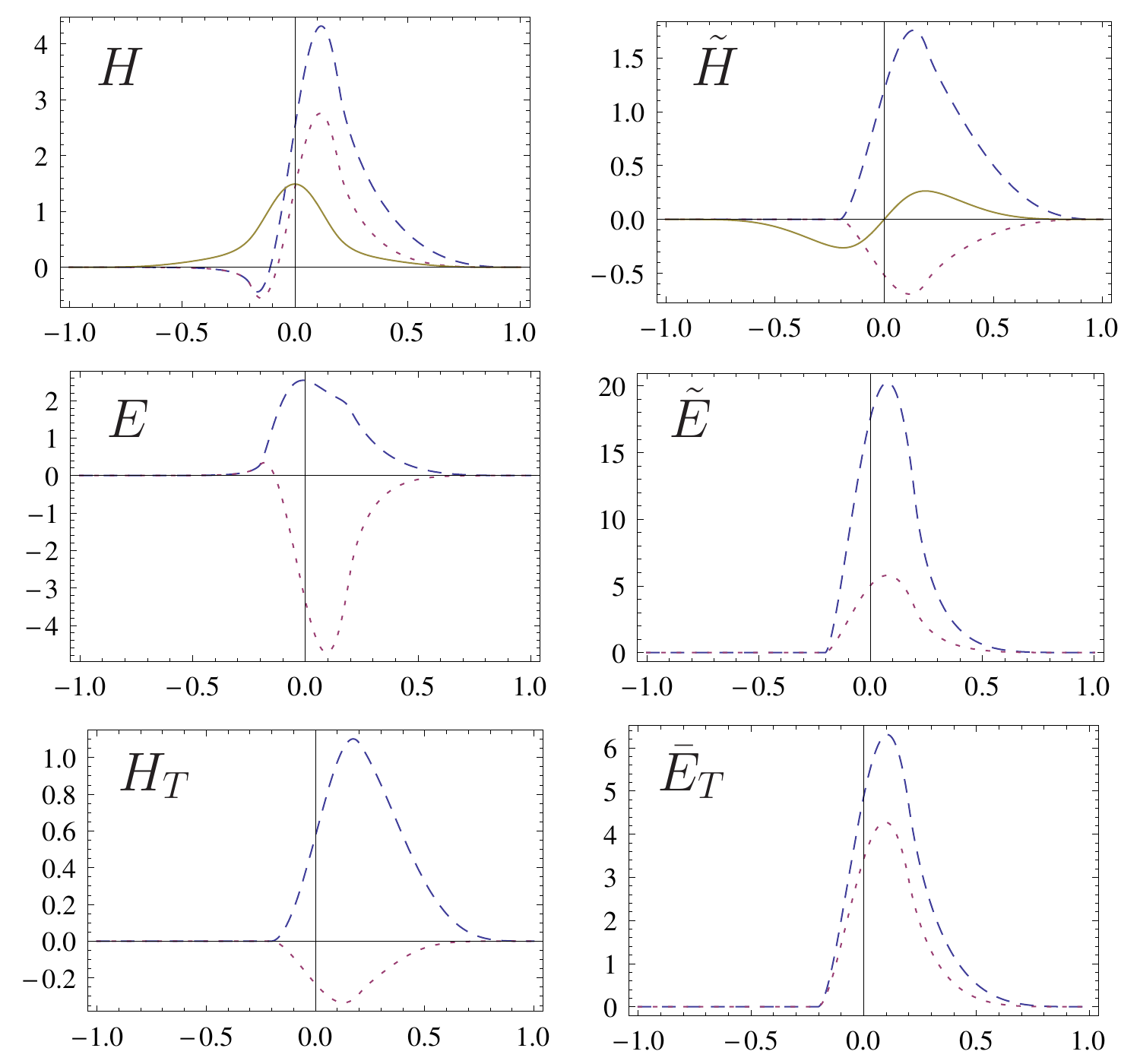}
\caption{Our model for the $x-$dependence of generalized parton distributions : $u-$quark (dashed lines), $d-$quark (dotted lines) and gluon (solid lines), for $\xi=0.2$ and $t= -0.15$ GeV$^2$.
}
\label{GPD}       
\end{figure}

We show on Fig. \ref{GPD} the $x-$dependence of the GPDs that we use here for $\xi=0.2$ and $t= -0.15$ GeV$^2$, which are characteristic values for our process in the present experimentally accessible domain. 

\section{The transverse amplitude.}
It is straightforward to show that the transverse amplitude vanishes at the leading twist level in the zero quark mass limit. For chiral-even GPDs, this comes from the colinear kinematics appropriate to the calculation of the leading twist coefficient function; for chiral-odd GPDs, this comes from the odd number of $\gamma$ matrices in the Dirac trace. This vanishing is related to the known results for the light meson electroproduction amplitudes \cite{transGPDno}.

To estimate the transverse amplitude, one thus needs to evaluate quark mass effects in the coefficient function and add the  part of the heavy meson DA which is proportional to the meson mass. 

With respect to the quark mass effects, it has been demonstrated \cite{Collins:1998rz} that hard-scattering factorization of meson leptoproduction \cite{fact3} is valid at leading twist with the inclusion of heavy quark masses in the hard amplitude. This proof is applicable independently of the relative sizes of the heavy quark masses and $Q$, and the size of the errors is a power of $\Lambda/ \sqrt{Q^2+M_D^2}$ when $\sqrt{Q^2+M_D^2}$ is the large scale. In our case, this means including the part $\frac{m_c}{k_c^2-m_c^2}$ in the off-shell heavy quark propagator (see the Feynman graph  on Fig. 1a) present in the leading twist coefficient function. 
We of course keep the term  $m_c^2$ in the denominator. 
Adding this part leads straightforwardly to a non-zero transverse amplitude when a chiral-odd transversity GPD is involved.
Including the mass term in the heavy meson DA may be a case of concern, since it is not the only twist 3 component and collinear factorization has not been proven for our process beyond the leading twist 2. We shall nonetheless make the assumption that our evaluation is legitimate. It turns out that this contribution, which we will trace by its proportionality to the meson mass $M_D$ is twice as large as the contribution coming form the inclusion the quark mass in the coefficient function.

In the Feynman gauge, the non-vanishing $m_c$ or $M_D$ dependent part of the Dirac trace in the hard scattering part depicted in Fig. 1a reads:\begin{eqnarray}
Tr[\sigma^{pi}\gamma^\nu (\hat p_D-M_D) \gamma^5 \gamma^{\nu'} \frac{\hat k_c+m_c}{k_c^2-m_c^2+ i \epsilon}(1+ \gamma^5)\hat \epsilon \frac{-g_{\nu \nu'}}{k_g^2+ i \epsilon} ] 
 = \frac{2 Q^2}{\xi'} \epsilon_\mu[ \epsilon^{\mu p i n} + i g^{\mu i}_\perp]  \frac{m_c-2M_D}{k_c^2-m_c^2+ i \epsilon} \frac{1}{k_g^2 + i \epsilon} \, , 
 \end{eqnarray} 
 where $k_c$ ($k_g$) is the heavy quark (gluon) momentum and $\epsilon$ the polarization vector of the $W-$boson. 
The fermionic trace vanishes for the diagram shown on Fig. 1b thanks to the identity $\gamma^\rho \sigma^{\alpha \beta}\gamma_\rho = 0$. The denominators of the propagators 
read:
\begin{eqnarray}
&&k_c^2 - m_c^2 +i\epsilon= \frac{Q^2 }{2 \xi'} (x+\xi-2\xi') - m_c^2 +i\epsilon =  \frac{Q^2+M_D^2 }{2 \xi} (x+\xi) - Q^2   - m_c^2 +i\epsilon \,,\\
 && k_g^2 +i\epsilon= \bar z [\bar z M_D^2 + \frac{Q^2 + M_D^2}{2 \xi} (x-\xi) +i\epsilon]\, . \nonumber
 \end{eqnarray}  
The transverse amplitude is then written  as ($\tau = 1-i2$):
\begin{eqnarray}
T_{T} & = &\frac{- i 2C_q \xi (2M_D-m_c)}{\sqrt 2 (Q^2+M_D^2)}  \bar{N}(p_{2}) \left[  {\mathcal{H}}_{T} i\sigma^{n\tau} +\tilde {\mathcal{H}}_{T}\frac{\Delta^{\tau}}{m_N^2} \right. \nonumber \\
&&  + {\mathcal E}_{T} \frac{\hat n \Delta ^{\tau}+2\xi  \gamma ^{\tau}}{2m_N} - \tilde {\mathcal E}_{T}\frac{\gamma ^{\tau}}{m_N}] N(p_{1}), 
\end{eqnarray}
with $C_q= \frac{2\pi}{3}C_F \alpha_s V_{dc}$, in terms of  transverse form factors that we define as  :
\begin{eqnarray}
{\cal F }_T=f_{D}\int \frac{\phi_D(z)dz}{\bar z}\hspace{-.1cm}\int \frac{F^d_T(x,\xi,t) dx }{(x-\xi+\beta \xi+i\epsilon) (x-\xi +\alpha \bar z+i\epsilon)},
\label{TFF}
 \end{eqnarray} 
where $F^d_T$ is any d-quark transversity GPD, $\alpha = \frac {2 \xi M_D^2}{Q^2+M_D^2}$, $\beta =  \frac {2 (M_D^2-m_c^2)}{Q^2+M_D^2}$. 

\begin{figure}
\includegraphics[width=0.8\textwidth]{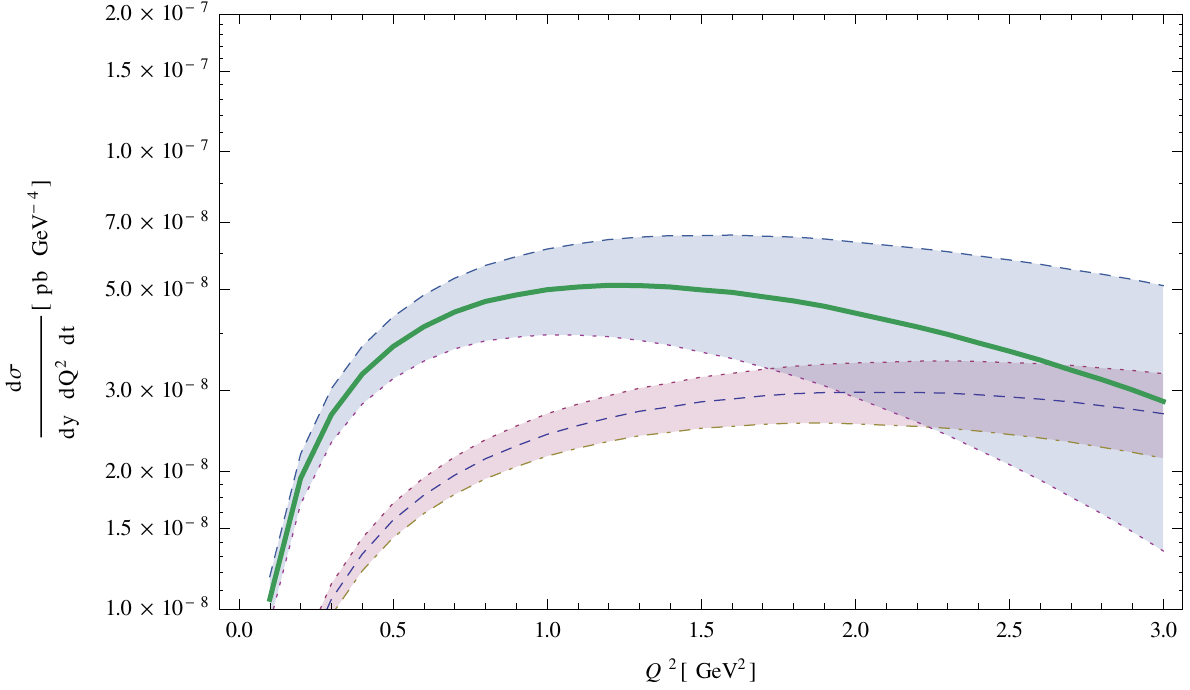}
\caption{The $Q^2$ dependence of the transverse contribution to the cross section $\frac{d\sigma(\nu N \to l^- N D^+)}{dy\, dQ^2\, dt}$ (in pb GeV$^{-4}$) for $y=0.7, \Delta_T = 0$  and $s=20$ GeV$^2$ for a proton (dashed curve and lower band) and neutron (solid curve and upper band) target.}
   \label{sigma{--}_dy_D+_on_proton_and_neutron}
\end{figure}

\begin{figure}
\includegraphics[width=0.8\textwidth]{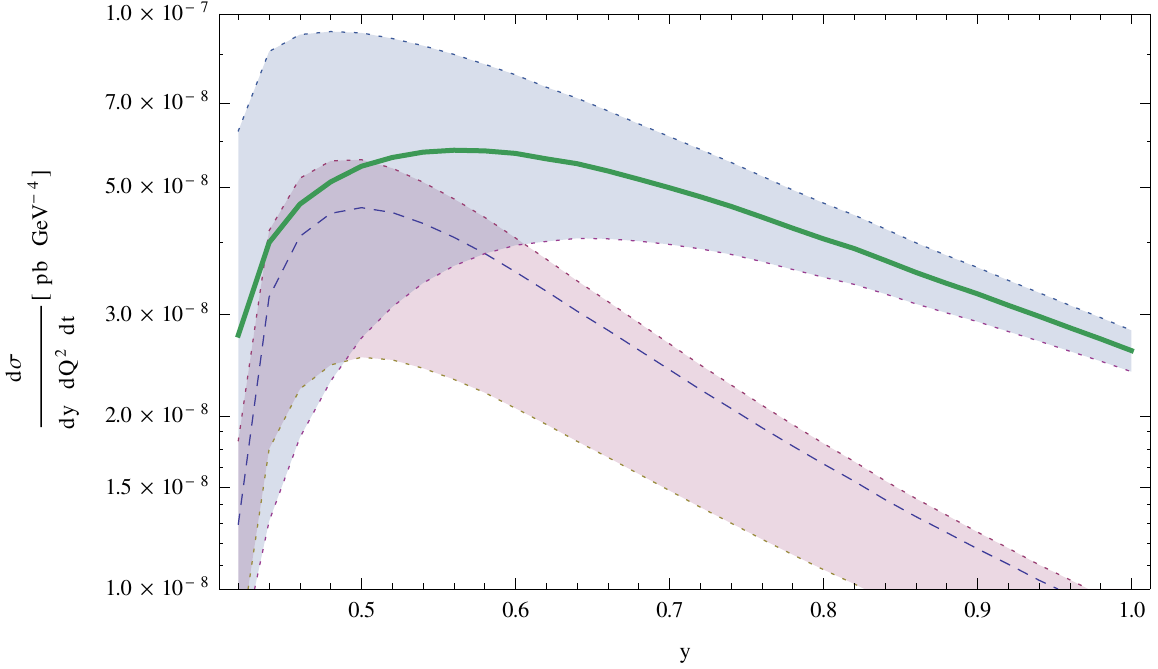}
\caption{The $y$ dependence of the transverse contribution to the cross section $\frac{d\sigma(\nu N \to l^- N D^+)}{dy\, dQ^2\, dt}$ (in pb GeV$^{-4}$) for $Q^2 = 1$ GeV$^2$, $\Delta_T = 0$  and $s=20$ GeV$^2$ for a proton (dashed curve) and neutron (solid curve) target.}
   \label{sigma{--}_dy_D+_on_proton_and_neutron_Q2_1_y}
\end{figure}

\begin{figure}
\includegraphics[width=0.95\textwidth]{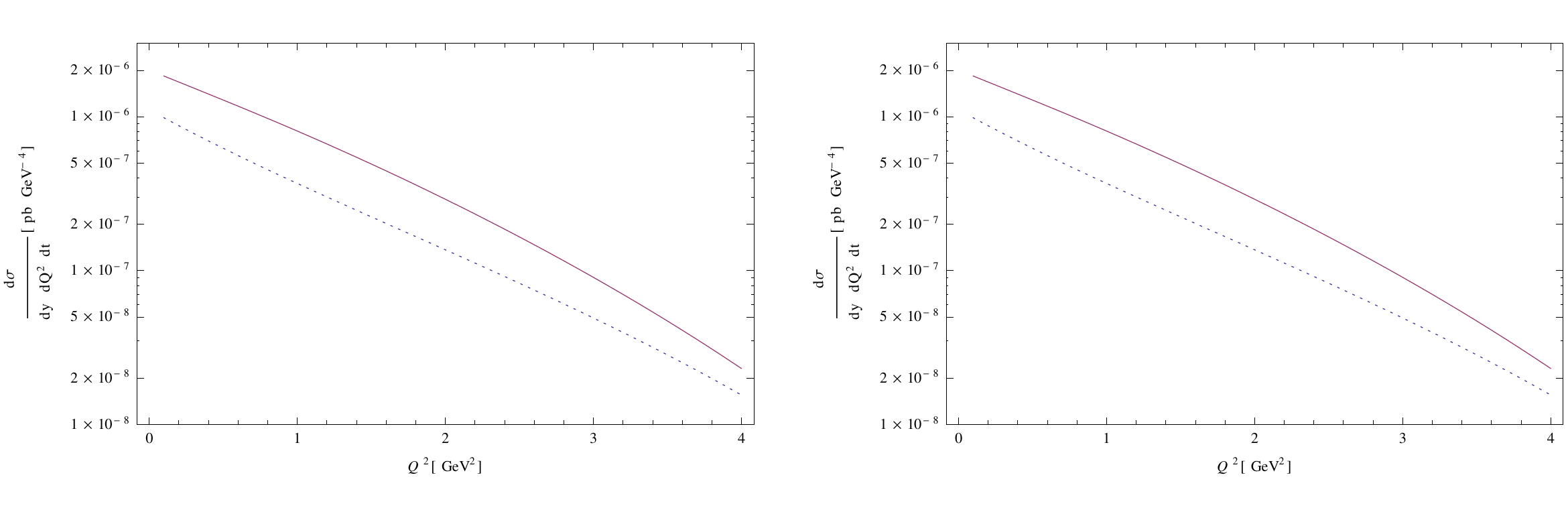}
\caption{The $Q^2$ dependence of the quark (dashed curve) contribution compared  to the total (quark and gluon, solid curve) longitudinal cross section $\frac{d\sigma(\nu N \to l^- N D^+)}{dy\, dQ^2\, dt}$ (in pb GeV$^{-4}$) for $D^+$ production  on a proton (left panel) and neutron (right panel) target for $y=0.7, \Delta_T = 0$  and $s=20$ GeV$^2$.}
   \label{sigma{00}_dy_D+_on_proton_neutron_quarks_gluons}
\end{figure}

\begin{figure}
\includegraphics[width=0.95\textwidth]{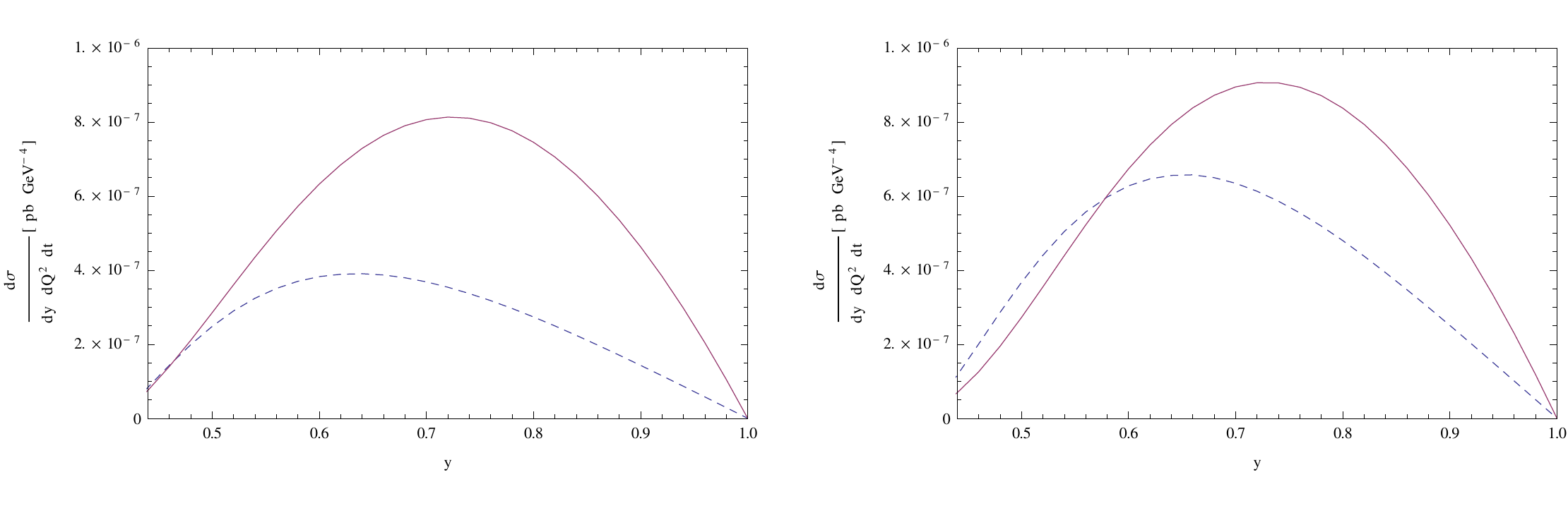}
\caption{The $y$ dependence  of the longitudinal contribution to the cross section $\frac{d\sigma(\nu N \to l^- N D^+)}{dy\, dQ^2\, dt}$ (in pb GeV$^{-4}$) for $Q^2 = 1$ GeV$^2$, $\Delta_T = 0$  and $s=20$ GeV$^2$ for a  proton (left panel) and neutron (right panel) target : total (quark and gluon, solid curve) and  quark only (dashed curve) contributions.}
   \label{sigma{00}_dy_D+_on_proton_neutron_quarks_gluons_y}
\end{figure}

 \section{The longitudinal amplitude.}
  When there is a change in the baryonic flavor, as in the reaction $\nu_l (k)n(p_1) \to l^- (k')D^0 (p_D)p'(p_2) $, the amplitude does not depend on gluon GPDs. In the other cases, namely $\nu_l (k)p(p_1) \to l^- (k')D^+ (p_D)p'(p_2) $ and $\nu_l (k)n(p_1) \to l^- (k')D^+ (p_D)n'(p_2) $, there is a gluonic contribution coming from the diagrams of Fig. 2.

 \subsection{The quark contribution.}
 The flavor sensitive electroweak vertex selects the $d \to c$ transition in the case of neutrino production, and the $\bar d \to \bar c$ transition in the case of antineutrino production. The fact that isospin relates the $d-$quark content of the neutron to the $u-$quark content of the proton, gives thus access to the $u-$quark GPDs in the proton when one scatters on a neutron. Moreover, reactions such as $\nu n \to l^- D^0 p$ give access to the neutron $\to$ proton GPDs ($H^{du}, E^{du}$...) which are related to the differences of GPDs in the proton through $H^{du}(x,\xi,t)=H^d(x,\xi,t)-H^u(x,\xi,t)$.
 The two Feynman diagrams of Fig. 1 contribute to the coefficient function. 
The contribution of diagram (a) to the hard amplitude involves one heavy quark propagator and thus has a part proportional  to $m_c M_D$.
The contribution of diagram (b) to the hard amplitude does not involve heavy quark propagator and the mass term in the meson DA does not contribute.
The chiral-odd GPDs do not contribute to the longitudinal amplitude since the coefficient function does not depend on any transverse vector. 

The vector and axial hard amplitudes (without the coupling constants) read:
\begin{eqnarray}
{\cal M}_H^V &=&   \left\{ \frac{Tr_a}{D^q_1 D^q_2} + \frac{Tr_b}{D^q_1 D^q_3}  \right\} \\
{\cal M}_H^5 &=&  \left\{ \frac{Tr_a^5}{D^q_1 D^q_2} + \frac{Tr_b^5}{D^q_1 D^q_3} \right\} \,,
\end{eqnarray}
where the propagators  are :
\begin{eqnarray}
D^q_1 &=& [ (x-\xi)p+ \bar z p_D ]^2+i\epsilon  = \bar z ^2 M_D^2+\bar z (x-\xi) \frac{Q^2+M_D^2}{2\xi }+i\epsilon\,,\nonumber\\
D^q_2 &=& [ (x+\xi)p+q ]^2-m_c^2+i\epsilon  =  \frac{Q^2+M_D^2}{2\xi} (x-\xi+\beta \xi+i\epsilon)\,,\nonumber\\
D^q_3 &=& [ 2 \xi p- \bar z p_D ]^2+i\epsilon  = \bar z ^2 M_D^2-\bar z (Q^2+M_D^2)+i\epsilon\,= -\bar z (Q^2+z M_D^2),
\end{eqnarray}
and the traces are :
\begin{eqnarray}
Tr_a &=& Tr_a^5 = 2\frac{Q^2+M_D^2}{\xi Q} \left[M_D^2-2M_Dm_c+(Q^2+M_D^2)\frac{x-\xi}{2\xi}\right] \,,\nonumber\\
Tr_b &=& Tr_b^5 = - 2\bar z Q^2 \frac{Q^2+M_D^2}{\xi Q}\,.
\end{eqnarray}

The quark contribution to the amplitude is a convolution of chiral-even GPDs $H^d(x,\xi,t), \tilde H^d(x,\xi,t), E^d(x,\xi,t)$ and $\tilde E^d(x,\xi,t)$ and reads:
\begin{eqnarray}
T_{L}^q  &=& \frac{- iC_q}{2Q} \bar{N}(p_{2}) \left[  {\mathcal{H}}_{L} \hat n - \tilde {\mathcal{H}}_{L} \hat n \gamma^5 
 + {\mathcal E}_{L} \frac{i\sigma^{n\Delta}}{2m_N} - \tilde {\mathcal E}_{L} \frac{\gamma ^5 \Delta.n}{2m_N}\right] N(p_{1}),
\end{eqnarray}
with the chiral-even form factors defined with a quite more intricate formula than  in Eq. (\ref{TFF}) by
\begin{eqnarray}
{\cal F }_L=f_{D}\int \frac{\phi_D(z)dz}{\bar z}\hspace{-.1cm}\int dx \frac{F^d(x,\xi,t)}{x-\xi +\alpha \bar z+i\epsilon}  \left[\frac{x-\xi +\gamma \xi }{x-\xi+\beta \xi+i\epsilon}+\frac{Q^2}{Q^2+zM_D^2}\right],
\label{LFF}
 \end{eqnarray} 
with $\gamma = \frac {2M_D(M_D-2m_c)}{Q^2+M_D^2}$, for any chiral even $d-$quark  GPD in the nucleon $F^d(x,\xi,t)$.

  \subsection{The gluonic contribution.}
The six Feynman diagrams of Fig. 2 contribute to the coefficient function when there is no charge exchange. Note that contrarily to the case of electroproduction of light pseudoscalar mesons, there is no $C-$parity argument to cancel the gluon contribution for the neutrino production of a pseudoscalar charmed meson.The  last three ones correspond to the  first three ones with the substitution $x\leftrightarrow -x$, and an overall minus sign for the axial case.
The contribution of diagrams (a)  and (d) to the hard amplitude does not involve any heavy quark propagator and the mass term in the meson DA does not contribute. 
The contribution of diagrams (b) and (e) to the hard amplitude involves two heavy quark propagators and they thus have a part proportional to $m_c^2$ as well as a part proportional to $m_c M_D$.
The contribution of diagrams  (c) and (f) involve one heavy quark propagator and they thus have a part proportional  to $m_c M_D$.
The transversity gluon GPDs do not contribute to the longitudinal amplitude since there there is no way to flip the helicity by two units when producing a (pseudo)scalar meson. This will not be the case for the production of a vector meson $D^*$.

The symmetric and antisymmetric hard amplitudes read:
\begin{eqnarray}
g_\perp^{ij} {\cal M}_H^S &=&   \left\{ \frac{Tr_a^S}{D_1 D_2} + \frac{Tr_b^S}{D_3 D_4} + \frac{Tr_c^S}{D_4 D_5}\right\}  + \left\{x\rightarrow  -x\right\} \\
i \epsilon_\perp^{ij}{\cal M}_H^{A} &=&  \left\{ \frac{Tr_a^A}{D_1 D_2} + \frac{Tr_b^A}{D_3 D_4} + \frac{Tr_c^A}{D_4 D_5} \right\}- \left\{x\rightarrow  -x\right\}
\end{eqnarray}
where the traces are:
\begin{eqnarray}
Tr_a^S &=&  \frac{2 \bar z }{Q} g_T^{ij}\left[z M_D^4 +Q^4+Q^2M_D^2(1+ z) - \frac{x+\xi}{2\xi} Q^2 (Q^2+M_D^2)\right]\,,\\
Tr_a^A &=&  \frac{2i \bar z \epsilon^{npij}}{Q} \left[z M_D^4 +Q^2M_D^2(1+\bar z) + \frac{x-\xi}{2\xi} Q^2 (Q^2+M_D^2)\right]\,,\\
Tr_b^S &=&   \frac{2(Q^2+M_D^2) }{Q} g_T^{ij}\left[- \frac{x+\xi}{2\xi} m_cM_D - z \frac{x-\xi}{2\xi}Q^2 +m_c^2+z \bar z M_D^2 \right]\,,\\
Tr_b^A &=&  \frac{2i  \epsilon^{npij}}{Q} \left[-z\frac{x-\xi}{2\xi} Q^4 +Q^2M_D^2(1+z)  (-z- \frac{x-\xi}{2\xi})+ M_D^4(z^2-z-1- \frac{x+\xi}{2\xi})\right.\nonumber \\
&+& \left. (M_D^2-m_c^2) (M_D^2-Q^2) +M_D(M_D+m_c)(Q^2+M_D^2)\frac{x+\xi}{2\xi} \right]\,,\\
Tr_c^S &=&  -\frac{Q^2+M_D^2}{\xi Q}g_T^{ij} \left[ (Q^2+M_D^2) \frac{x^2-\xi^2}{2\xi} +2 z M_D^2(\xi z+x) -M_D(m_c+M_D)(x-\xi+2\xi z)         \right] \,,\\
Tr_c^A &=&   \frac{2i  \epsilon^{npij}}{Q} \left[(z^2-1)Q^2M_D^2 +(\bar z M_D^2-(Q^2+M_D^2)\frac{x+\xi}{2\xi})((1+ z) M_D^2+(Q^2+M_D^2)\frac{x-\xi}{2\xi})\right.\nonumber\\ 
&+& \left.M_D(m_c+M_D)[(Q^2+M_D^2)\frac{x+\xi}{2\xi} +\bar z(Q^2-M_D^2)]\right]\,,
\end{eqnarray}
and the denominators read
\begin{eqnarray}
D_1 &=& \bar z [  -z M_D^2 -Q^2 +i \varepsilon] \,,\\
D_2 &=& \bar z [ \bar z M_D^2 +\frac{x-\xi}{2\xi}(Q^2+M_D^2) +i \varepsilon] =  \bar z \frac{Q^2+M_D^2}{2\xi} (x-\xi+\alpha\bar z +i\epsilon)\,,\\
D_3 &=& -z Q^2-z \bar z M_D^2- m_c^2 +i \varepsilon = -z(Q^2+M_D^2) +z^2M_D^2-m_c^2 +i\epsilon \,,\\
D_4 &=& z^2 M_D^2 -m_c^2+\frac{z(x-\xi)}{2\xi}(Q^2+M_D^2) +i \varepsilon\,,\\
D_5 &=& \bar z [ \bar z M_D^2 -\frac{x+\xi}{2\xi}(Q^2+M_D^2) +i \varepsilon]  = \bar z \frac{Q^2+M_D^2}{2\xi} (-x-\xi+\alpha\bar z +i\epsilon)\,.
\end{eqnarray}
The gluonic contribution to the amplitude thus reads:
\begin{eqnarray}
T_L^g &=&  \frac{ i C_g}{2} \int_{-1}^{1}dx \frac{- 1}{(x+\xi-i\epsilon)(x-\xi+i\epsilon)} \int_0^1 dz f_D \phi_D(z) \cdot \nonumber \\
&& \left[  \bar{N}(p_{2})[H^g\hat n +E^g\frac{i\sigma^{n\Delta}}{2m} ]N(p_{1}) {\cal M}_H^S+\bar{N}(p_{2})[{\tilde H}^g \hat n \gamma^5+{\tilde E}^g\frac{\gamma^5n.\Delta}{2m} ]N(p_{1}) {\cal M}_H^{A}  \right] \, \\
&\equiv&  \frac{- i C_g}{2Q}  \bar{N}(p_{2}) \left[ {\cal H}^g\hat n +{\cal E}^g\frac{i\sigma^{n\Delta}}{2m} +{\tilde {\cal H}}^g \hat n \gamma^5+{\tilde {\cal E}}^g\frac{\gamma^5n.\Delta}{2m} \right] N(p_{1}) \,, 
\end{eqnarray}
where  the last line defines the gluonic form factors ${\cal H}^g$, $\tilde {\cal H}^g$, ${\cal E}^g$, $\tilde {\cal E}^g$ and  $C_g= T_f\frac{\pi}{3} \alpha_s V_{dc}$ with $T_f=\frac{1}{2}$ and the factor $\frac{- 1}{(x+\xi-i\epsilon)(x-\xi+i\epsilon)}$ comes from the conversion of the strength tensor to the  gluon field. Note that there is no singularity in the integral over $z$ if the DA vanishes like $z \bar z$ at the limits of integration.

\begin{figure}
\includegraphics[width=0.8\textwidth]{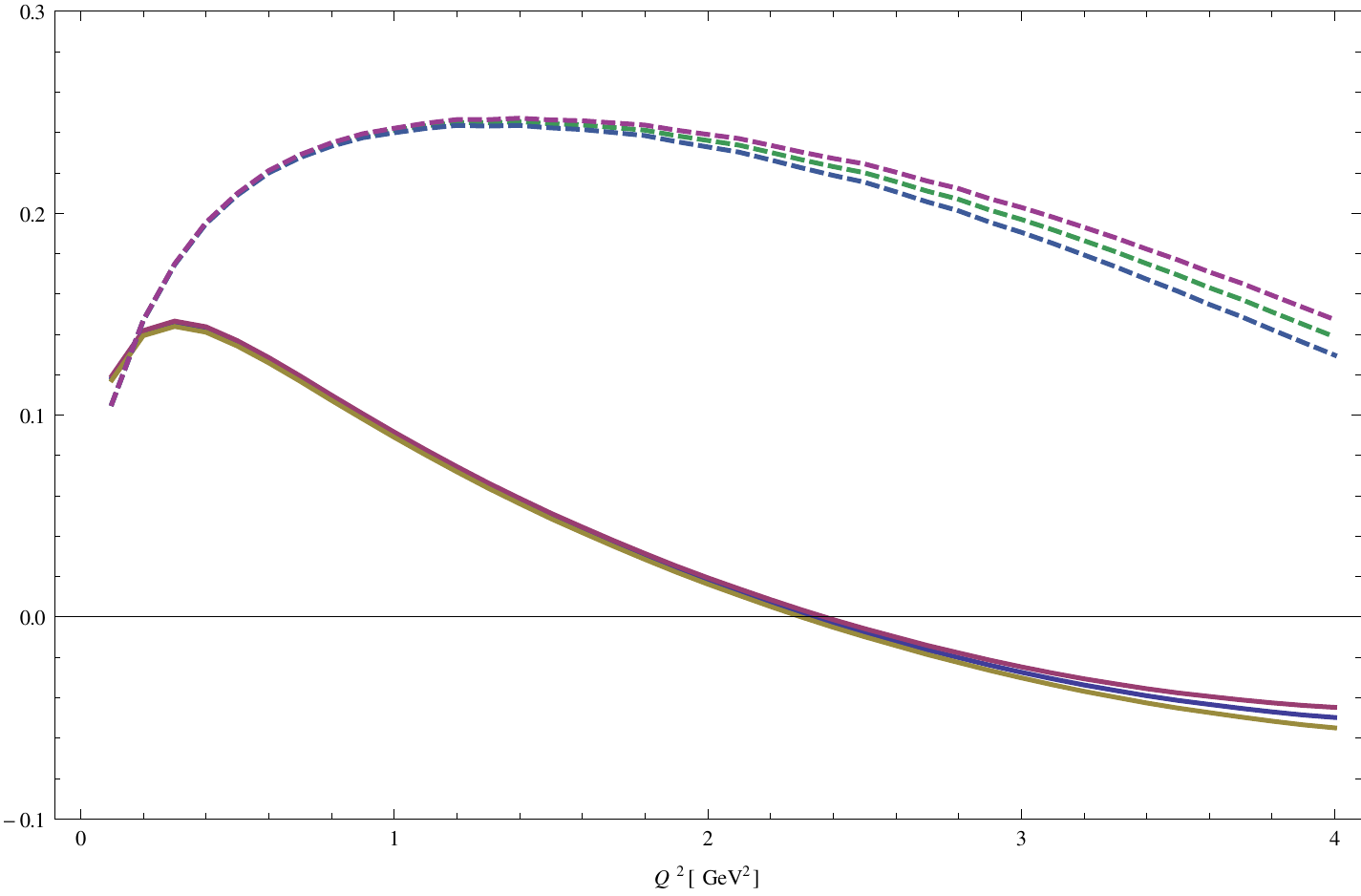}
\caption{The $Q^2$ dependence of the  $<cos~ \varphi> $ (solid curves) and $<sin~ \varphi> $ (dashed curves) moments normalized by the total cross section, as defined  in Eq.\ref{moments}, for  $\Delta_T = 0.5$ GeV, $y=0.7$ and $s=20$ GeV$^2$. The  three curves correspond to the three models explained in the text, and quantify the theoretical uncertainty of our estimates.}
   \label{cosPhi-proton}
\end{figure}

\begin{figure}
\includegraphics[width=0.8\textwidth]{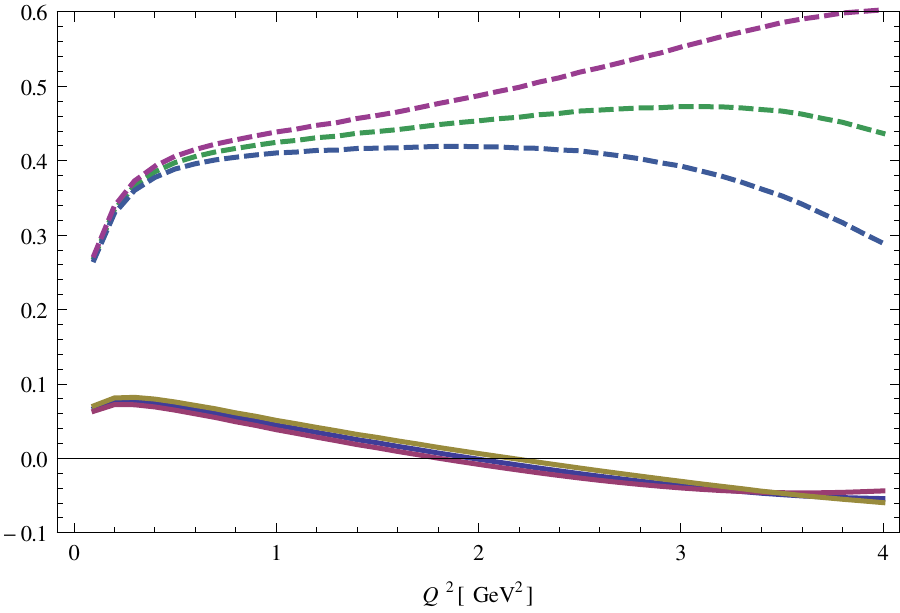}
\caption{The same   $Q^2$ dependence of the  $<cos~ \varphi> $ (solid curves) and $<sin~ \varphi> $ (dashed curves) moments as in Fig. \ref{cosPhi-proton}  but for a neutron target.}
   \label{cosPhi-neutron}
\end{figure}

 \section{Observables}
  The differential cross section for neutrino production of a pseudoscalar charmed boson is written as \cite{Arens}:
 \begin{eqnarray}
\label{cs}
\frac{d^4\sigma(\nu N\to l^- N'D)}{dy\, dQ^2\, dt\,  d\varphi}
 = \bar\Gamma
\Bigl\{ ~\frac{1+ \sqrt{1-\varepsilon^2}}{2} \sigma_{- -}&+&\varepsilon\sigma_{00}+  \sqrt{\varepsilon}(\sqrt{1+\varepsilon}+\sqrt{1-\varepsilon} )(\cos\varphi\
{\rm Re}\sigma_{- 0} + \sin\varphi\
 {\rm Im}\sigma_{- 0} )\ \Bigr\},
\end{eqnarray}
with $y= \frac{p \cdot q}{p\cdot k}$ , $Q^2 = x_B y (s-m^2)$, $\varepsilon \approx \frac{1-y}{1-y+y^2/2}$ and
\begin{equation}
\bar \Gamma = \frac{G_F^2}{(2 \pi)^4} \frac{1}{32y} \frac{1}{\sqrt{ 1+4x_B^2m_N^2/Q^2}}\frac{1}{(s-m_N^2)^2} \frac{Q^2}{1-\epsilon}\,, \nonumber
\end{equation}
where the ``cross-sections'' $\sigma_{lm}=\epsilon^{* \mu}_l W_{\mu \nu} \epsilon^\nu_m$ are product of  amplitudes for the process $ W(\epsilon_l) N\to D N' $, averaged  (summed) over the initial (final) hadron polarizations. Integrating over $\varphi$ yields the differential cross section :
 \begin{eqnarray}
\label{csphi}
\frac{d\sigma(\nu N\to l^- N'D)}{dy\, dQ^2\, dt}
 = 2 \pi \bar\Gamma
\Bigl\{ ~\frac{1+ \sqrt{1-\varepsilon^2}}{2} \sigma_{- -}&+&\varepsilon\sigma_{00} \Bigr\}.
\end{eqnarray}

We now calculate from $T_{L} $ and $T_{T} $ the  quantities $\sigma_{00}$, $\sigma_{--}$ and  $\sigma_{-0}$.
 The longitudinal cross section $\sigma_{00}$ is straightforwardly obtained by squaring the sum of the amplitudes $T^q_{L}+ T^g_{L}$; at zeroth order in $\Delta_T$, it reads  :
\begin{eqnarray}
\sigma_{00} =    \frac{1} { Q^2}\biggl\{ && [\, |C_q{\mathcal{H}}_{L} + C_g{\mathcal{H}}_{g}|^2 + |C_q\tilde{\mathcal{H}}_{L} - C_g\tilde{\mathcal{H}}_{g}|^2 ] (1-\xi^2) +\frac{\xi^4}{1-\xi^2} [\, |C_q\tilde{\mathcal{E}}_{L}-C_g\tilde{\mathcal{E}}_{g} |^2 +  |C_q {\mathcal{E}}_{L}+C_g {\mathcal{E}}_{g} |^2]  \nonumber  \\
  &&  -2 \xi^2 {\mathcal R}e  [C_q{\mathcal{H}}_{L} + C_g{\mathcal{H}}_{g}] [C_q {\mathcal{E}}^*_{L}+C_g {\mathcal{E}}^*_{g}]  -2 \xi^2 {\mathcal R}e  [C_q\tilde{\mathcal{H}}_{L} - C_g\tilde{\mathcal{H}}_{g}] [C_q \tilde{\mathcal{E}}^*_{L}-C_g \tilde{\mathcal{E}}^*_{g}] \biggr\} .\, 
\end{eqnarray}
 At zeroth order in $\Delta_T$, $\sigma_{--}$ reads:
\begin{eqnarray}
\sigma_{--} =   \frac{16\xi^2 C_q^2 (m_c-2M_D)^2}{(Q^2+M_D^2)^2}\biggl\{(1-\xi^2)|{\mathcal{H}_T}|^2  + \frac{\xi^2}{1-\xi^2} | { {\mathcal{E'}}_T}|^2 -2\xi \mathcal{R}e [ \mathcal{H}_T { {\mathcal{E'}}_T^{ *}}]\biggr\} \,, 
\end{eqnarray}
where we denote ${{\mathcal{E'}}_T}=\xi{\mathcal{E}}_T-{\tilde {\mathcal{E}}_T}$.

\begin{figure}
\includegraphics[width=0.8\textwidth]{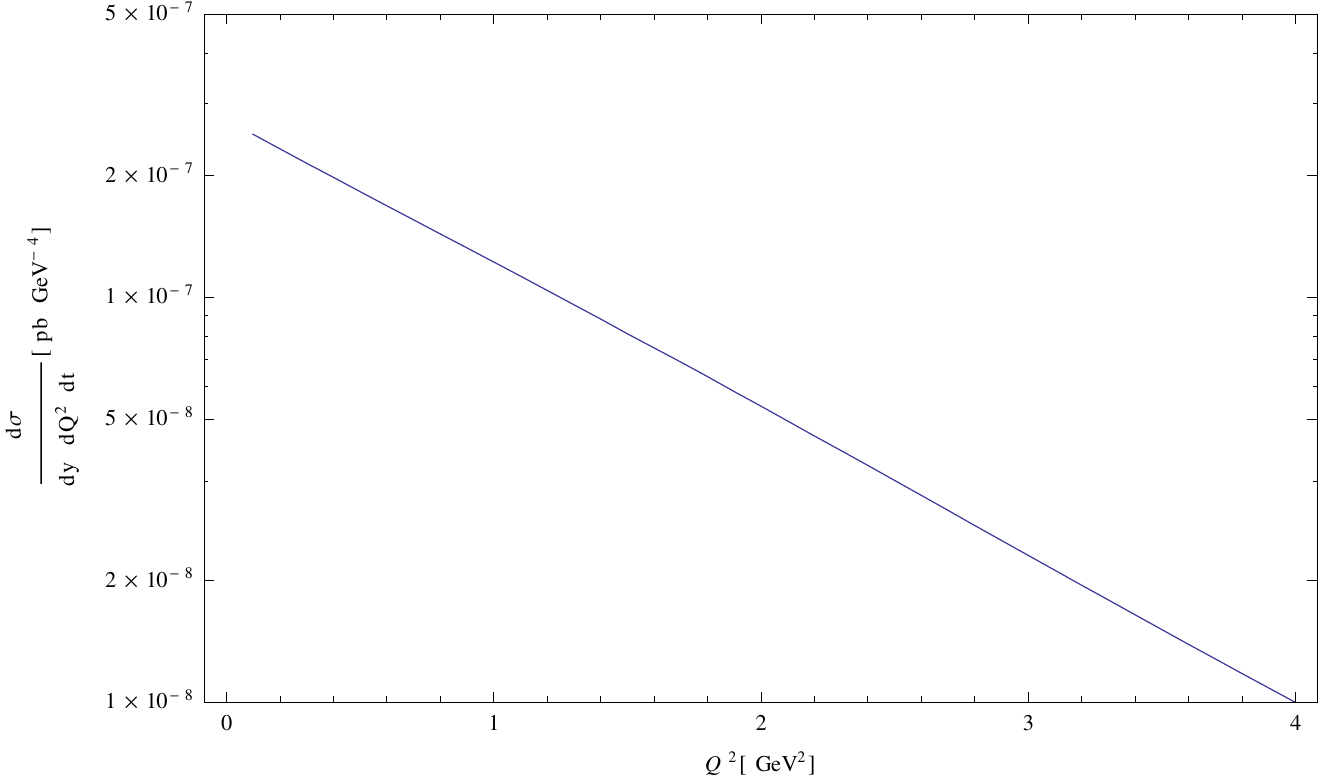}
\caption{The $Q^2$ dependence of the longitudinal contribution to the cross section $\frac{d\sigma(\nu n \to l^- p D^0)}{dy\, dQ^2\, dt}$ (in pb GeV$^{-4}$) for $y=0.7, \Delta_T = 0$  and $s=20$ GeV$^2$.}
   \label{sigma{00}_dy_D0}
\end{figure}

\begin{figure}
\includegraphics[width=0.8\textwidth]{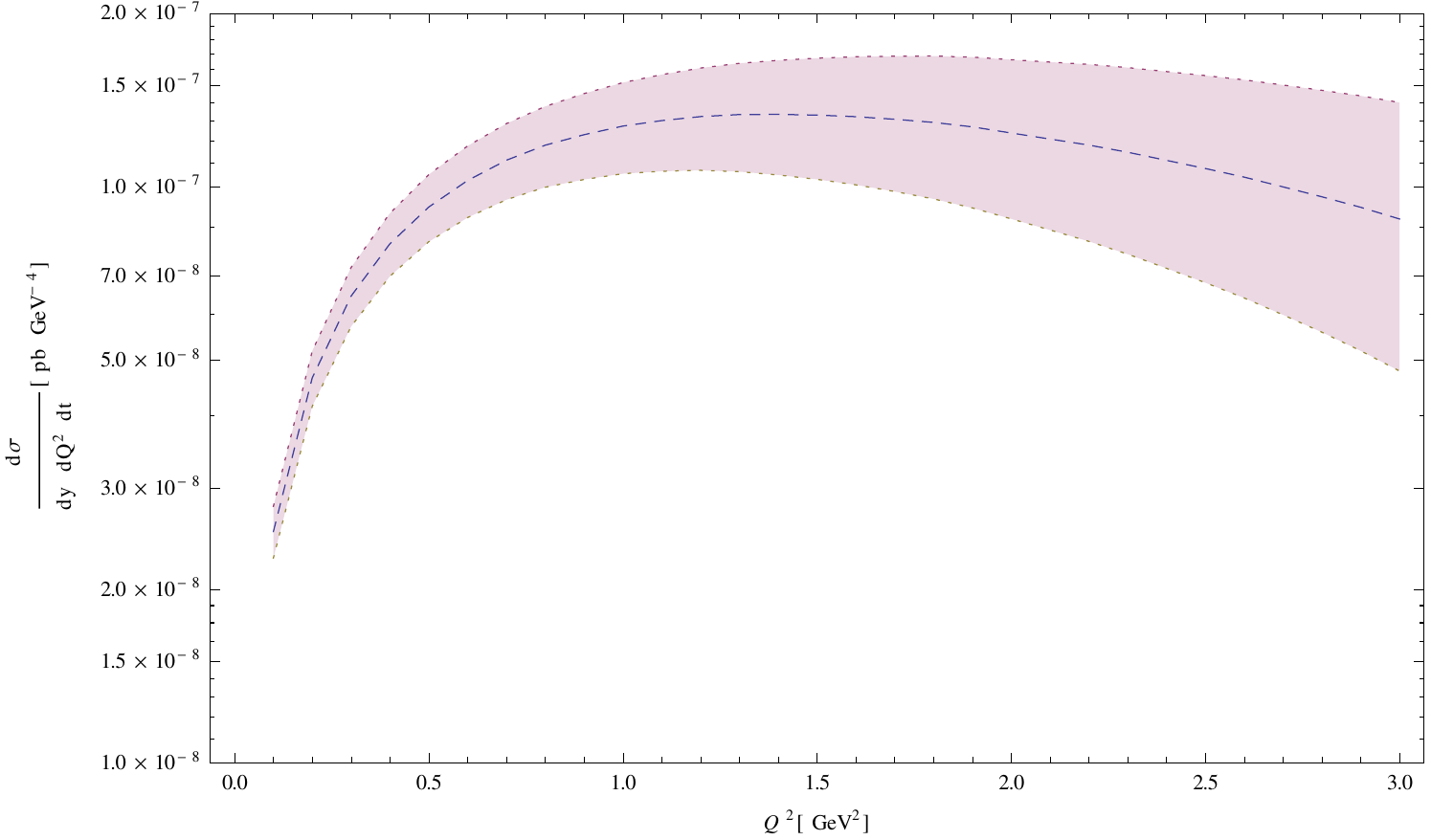}
\caption{The $Q^2$ dependence of the transverse contribution to the cross section $\frac{d\sigma(\nu n \to l^- p D^0)}{dy\, dQ^2\, dt }$ (in pb GeV$^{-4}$) for $y=0.7, \Delta_T = 0$  and $s=20$ GeV$^2$. The  band corresponds to the three models explained in the text.}
   \label{sigma{--}_dy_nnu_to_plD0}
\end{figure}

\begin{figure}
\includegraphics[width=0.8\textwidth]{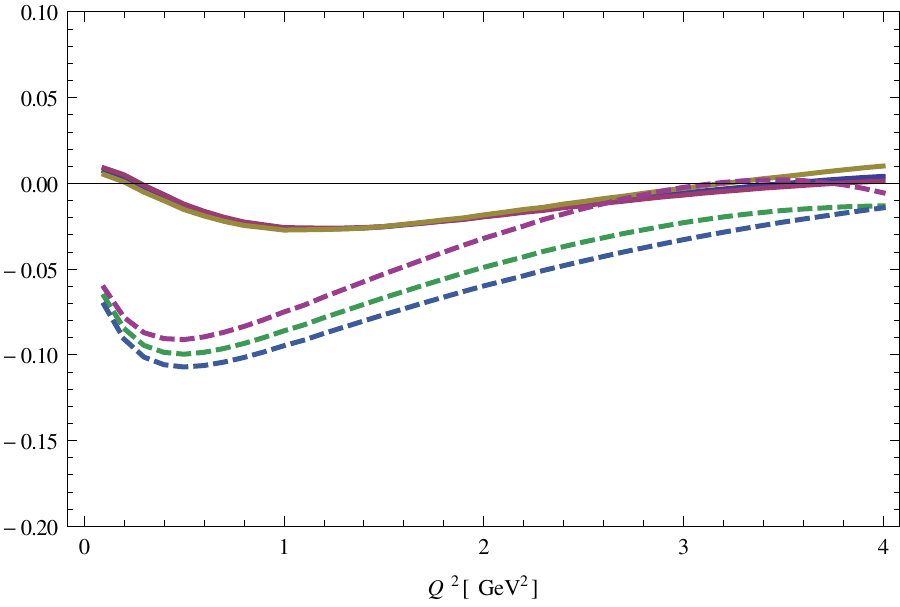}
\caption{  The $Q^2$ dependence of the $<cos~ \varphi> $ (solid curves) and $<sin~ \varphi> $ (dashed curves) moments normalized by the total cross section, as defined  in Eq.\ref{moments}, for  the reaction $\nu n \to l^- p D^0$ when $\Delta_T = 0.5$ GeV, $y=0.7$ and $s=20$ GeV$^2$ . The  three curves correspond to the three models explained in the text.}
   \label{CosPhi_SinPhi_D0}
\end{figure}

The  interference cross section   $\sigma_{-0}$ vanishes at zeroth order in $\Delta_T$. The first non-vanishing contribution is linear  in $\Delta_T/m$ and reads (with $\lambda = \tau^* = 1+i2$):
\begin{eqnarray}
\sigma_{-0} &=& 
\frac{ \xi \sqrt 2 C_q}{m}  \frac{ 2M_D- m_c}{Q(Q^2+M_D^2)} \,
\biggl\{-i \mathcal{H}_T^{*} [C_q\tilde{\mathcal{E}}_{L}-C_g\tilde{\mathcal{E}}_{g}]  \xi \epsilon^{pn\Delta  \lambda }  + i {{\mathcal{E'}}_T^{*}}\epsilon^{p n \Delta \lambda } [C_q\tilde {\mathcal{H}}_{L}-C_g \tilde {\mathcal{H}}_{g}] \nonumber \\
& + & 2 \mathcal{\tilde H}_T^{*} \Delta^\lambda \{C_q{\mathcal{H}}_{L}+C_g{\mathcal{H}}_{g} -\frac{\xi^2}{1-\xi^2} [C_q{\mathcal{E}}_{L}+C_g{\mathcal{E}}_{g}]\}  + \mathcal{E}_T ^{*} \Delta^\lambda \{(1-\xi^2) [C_q{\mathcal{H}}_{L}+C_g{\mathcal{H}}_{g}] -\xi^2 [C_q{\mathcal{E}}_{L}+C_g{\mathcal{E}}_{g}]\}  \nonumber \\
&-&   \mathcal{ H}_T^{*} \Delta^\lambda [C_q{\mathcal{E}}_{L}+C_g{\mathcal{E}}_{g}]   + {{\mathcal{E'}}_T^{*}} \Delta^\lambda \xi [C_q{\mathcal{H}}_{L}+C_g{\mathcal{H}}_{g} +C_q{\mathcal{E}}_{L}+C_g{\mathcal{E}}_{g} ]  \biggr\}.  \
\end{eqnarray}

As discussed in \cite{PS} the transversity GPDs are best accessed through the moments   $<cos~ \varphi>$ and $<sin ~\varphi>$ defined as

     \begin{eqnarray}
  <cos ~\varphi>&=&\frac{\int cos ~\varphi ~d\varphi ~d^4\sigma}{\int d\varphi ~d^4\sigma} \approx K_\epsilon\, \frac{{\cal R}e \sigma_{- 0}}{\sigma_{0 0}+K_\epsilon^2 \sigma_{--}}  \,, \nonumber \\
   <sin~ \varphi>&=&\frac{\int sin ~\varphi ~d\varphi ~d^4\sigma}{\int d\varphi ~d^4\sigma}\approx K_\epsilon \frac{{\cal I}m \sigma_{- 0}}{\sigma_{0 0}+K_\epsilon^2 \sigma_{--}}  \,,
   \label{moments}
   \end{eqnarray} 
   with $K_\epsilon =\frac{\sqrt{1+\varepsilon}+\sqrt{1-\varepsilon} }{2 \sqrt{\epsilon} }$. 

\subsection{$\nu N \to l^- D^+ N$}

Let us now estimate various cross sections which may be accessed with a neutrino beam on a nucleus. Firstly, 
\begin{eqnarray}
\nu_l (k)p(p_1) &\to& l^- (k')D^+ (p_D)p'(p_2) \,,\nonumber\\
\nu_l (k)n(p_1) &\to& l^- (k')D^+ (p_D)n'(p_2) \,, \nonumber
\end{eqnarray}
allow both quark and gluon GPDs to contribute. Neglecting the strange content of nucleons leads to selecting $d$ quarks in the nucleon, thus accessing the $d$ (resp. $u$) quark GPDs in the proton for the scattering on a proton  (resp. neutron) target, after using isospin relation between the proton and neutron.
The transverse contribution is plotted in Fig. \ref{sigma{--}_dy_D+_on_proton_and_neutron} as a function of $Q^2$  for $y=0.7$ and $\Delta_T=0$. The dependence on $y$ is shown on Fig. \ref{sigma{--}_dy_D+_on_proton_and_neutron_Q2_1_y} for $Q^2=1$ GeV$^2$ and $\Delta_T=0$.  The cross section is reasonably flat in $y$ and $Q^2$ so that an integration over the regions $0.45 < y <1$ and $0.5 < Q^2 < 3$ GeV$^2$ does not require much care.

The longitudinal cross sections dominate the transverse ones, mostly because of the larger values of the chiral-even GPDs, and specifically of the gluonic ones.
The relative importance of quark and gluon contributions to the longitudinal cross sections are shown on Fig. \ref{sigma{00}_dy_D+_on_proton_neutron_quarks_gluons} as a function of $Q^2$ for a specific set of kinematical variables. The $y$ dependence is displayed on Fig. \ref{sigma{00}_dy_D+_on_proton_neutron_quarks_gluons_y}. The longitudinal cross section vanishes as $y\to 1$ as is obvious from Eq. (\ref{cs}).

Access to the interference term $\sigma_{-0}$ needs an harmonic analysis of the cross section. This allows access to the transversity GPDs in a linear way but requires to consider $\Delta_T \ne 0$ kinematics.
We show on Fig. \ref{cosPhi-proton} the $<cos\varphi>$  and $<sin\varphi>$ moments for the proton and on Fig. \ref{cosPhi-neutron} for the neutron target, for the kinematical point defined as $y=0.7, \Delta_T = 0.5$ GeV and $s=20$ GeV$^2$. 

\subsection{$\nu n \to l^- D^0 p$}

The reaction
\begin{eqnarray}
\nu_l (k)n(p_1) &\to& l^- (k')D^0 (p_D)p(p_2) \,, \nonumber
\end{eqnarray}
does not benefit from gluon GPDs contributions, but only from the flavor changing $F_{du} (x, \xi, t) =F_d (x, \xi, t)-F_u(x, \xi, t)$ GPDs ($F$ denotes here any GPD). We show on Fig. \ref{sigma{00}_dy_D0} the longitudinal cross section and on Fig. \ref{sigma{--}_dy_nnu_to_plD0} the transverse one. This transverse contribution is noteworthy of the same order of magnitude as the longitudinal one and even dominates for $y$ large enough. Accessing the chiral-odd transversity GPDs indeed seems feasible in this reaction.

We  plot on Fig. \ref{CosPhi_SinPhi_D0} the $<cos~\varphi>$ and $<sin~\varphi>$ moments defined in Eq. (\ref{moments}) for this reaction. These moments are definitely smaller than in the $D^+$ production case.

\subsection{Antineutrino cross sections :  $\bar \nu p \to l^+ \bar D^0 n$ and $\bar \nu N \to l^+ \bar D^- N$}
For completeness, let us now present some results for the antineutrino case. Although smaller than the neutrino flux, the antineutrino flux is always sizable, as discussed recently in \cite{Ren:2017xov}.

\begin{figure}
\includegraphics[width=0.8\textwidth]{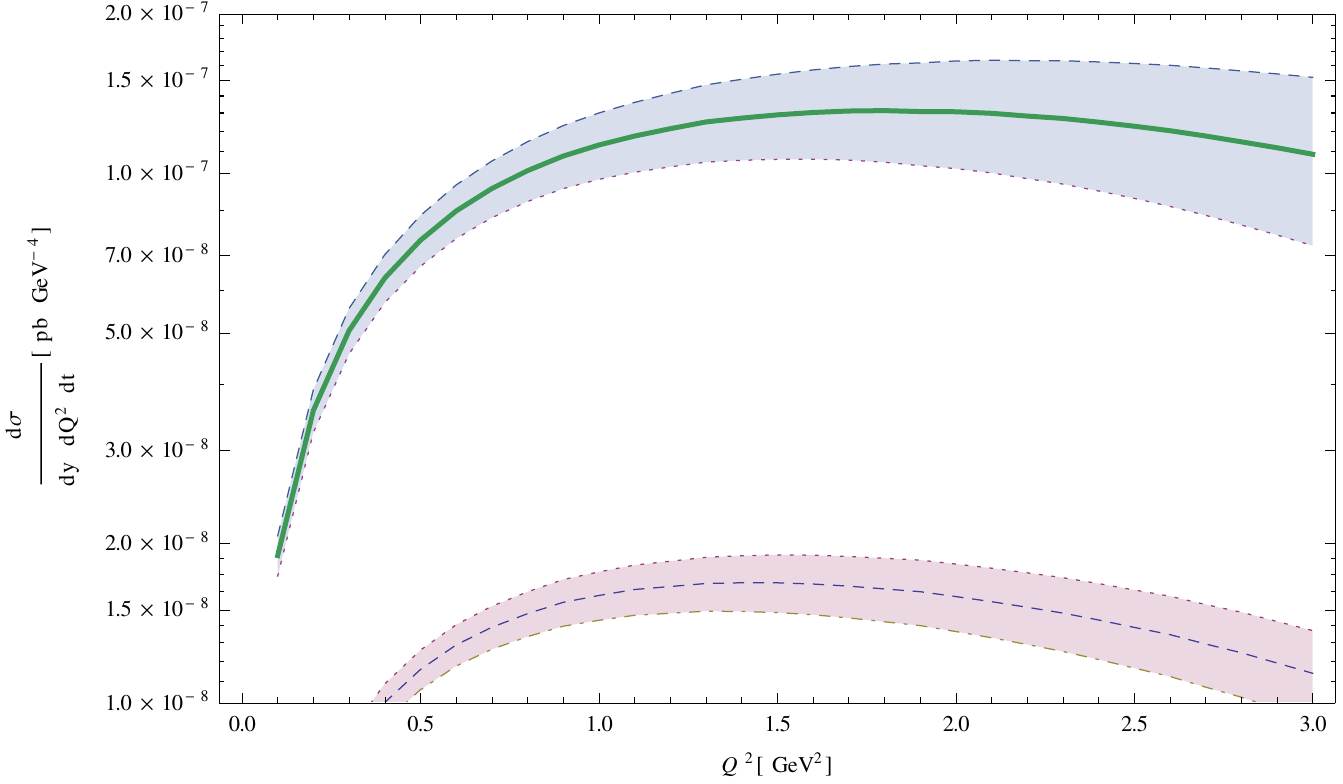}
\caption{The $Q^2$ dependence of the transverse contribution to the cross section $\frac{d\sigma(\bar \nu N \to l^+ N \bar D^-)}{dy\, dQ^2\, dt}$ on a proton (dashed curve) or neutron (solid curve) target (in pb GeV$^{-4}$) for $y=0.7, \Delta_T = 0$  and $s=20$ GeV$^2$. The  three curves correspond to the three models explained in the text, and quantify the theoretical uncertainty of our estimates.}
   \label{anti_sigma{--}_dy_D-_on_proton_and_neutron}
\end{figure}

\begin{figure}
\includegraphics[width=0.8\textwidth]{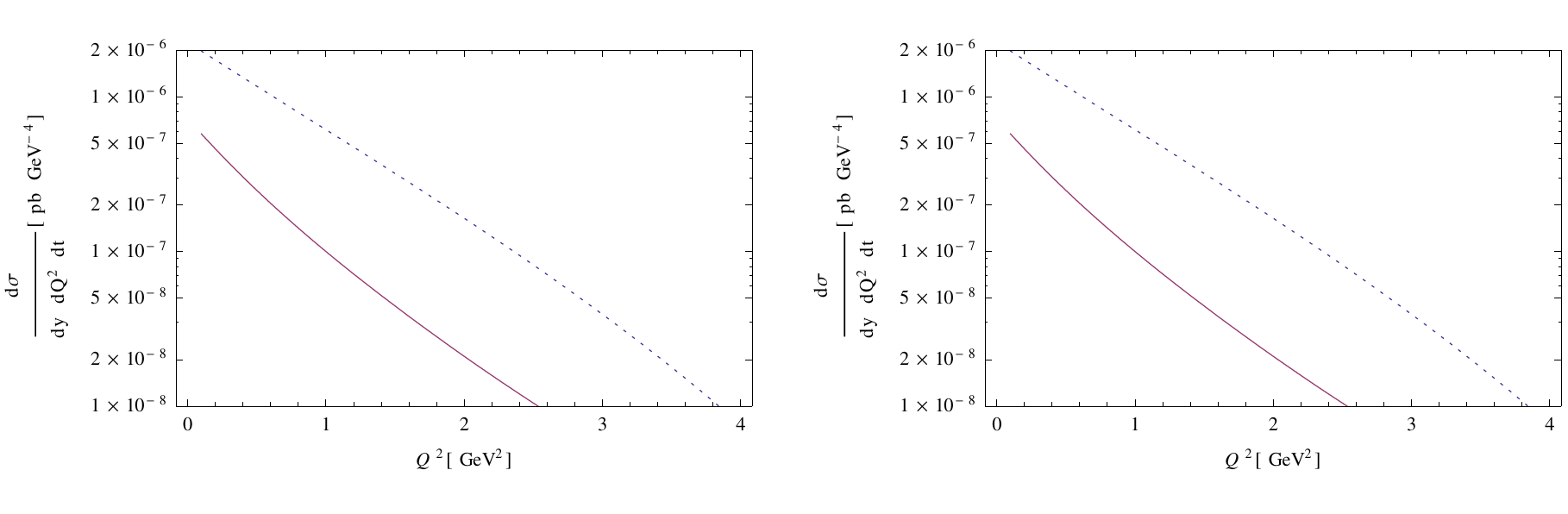}
\caption{The $Q^2$ dependence of the longitudinal contribution to the cross section $\frac{d\sigma(\bar \nu N \to l^+ N \bar D^-)}{dy\, dQ^2\, dt}$ on a proton (left panel) or neutron (right panel) target (in pb GeV$^{-4}$) for $y=0.7, \Delta_T = 0$  and $s=20$ GeV$^2$. The total (solid curve) is small with respect to the quark contribution (dashed curve), showing a large cancellation of quark and gluon contributions.}
   \label{anti_sigma{00}_dy_D-_on_proton_neutron_quarks_gluons}
\end{figure}

\begin{figure}
\includegraphics[width=0.8\textwidth]{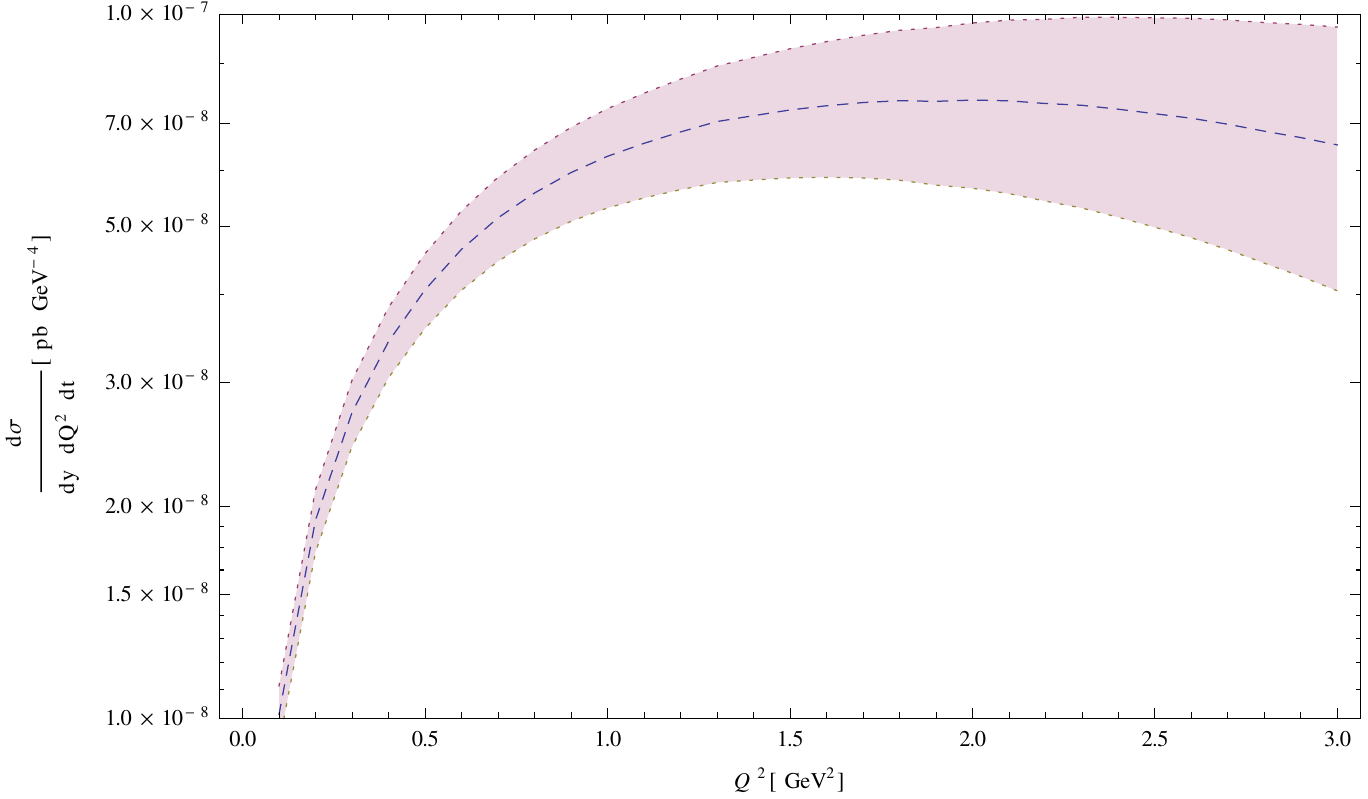}
\caption{The $Q^2$ dependence of the transverse contribution to the cross section $\frac{d\sigma(\bar \nu p \to l^+ n \bar D^0)}{dy\, dQ^2\, dt}$ (in pb GeV$^{-4}$) for $y=0.7, \Delta_T = 0$  and $s=20$ GeV$^2$. The  three curves correspond to the three models explained in the text, and quantify the theoretical uncertainty of our estimates.}
   \label{anti_sigma{--}_D0}
\end{figure}

\begin{figure}
\includegraphics[width=0.8\textwidth]{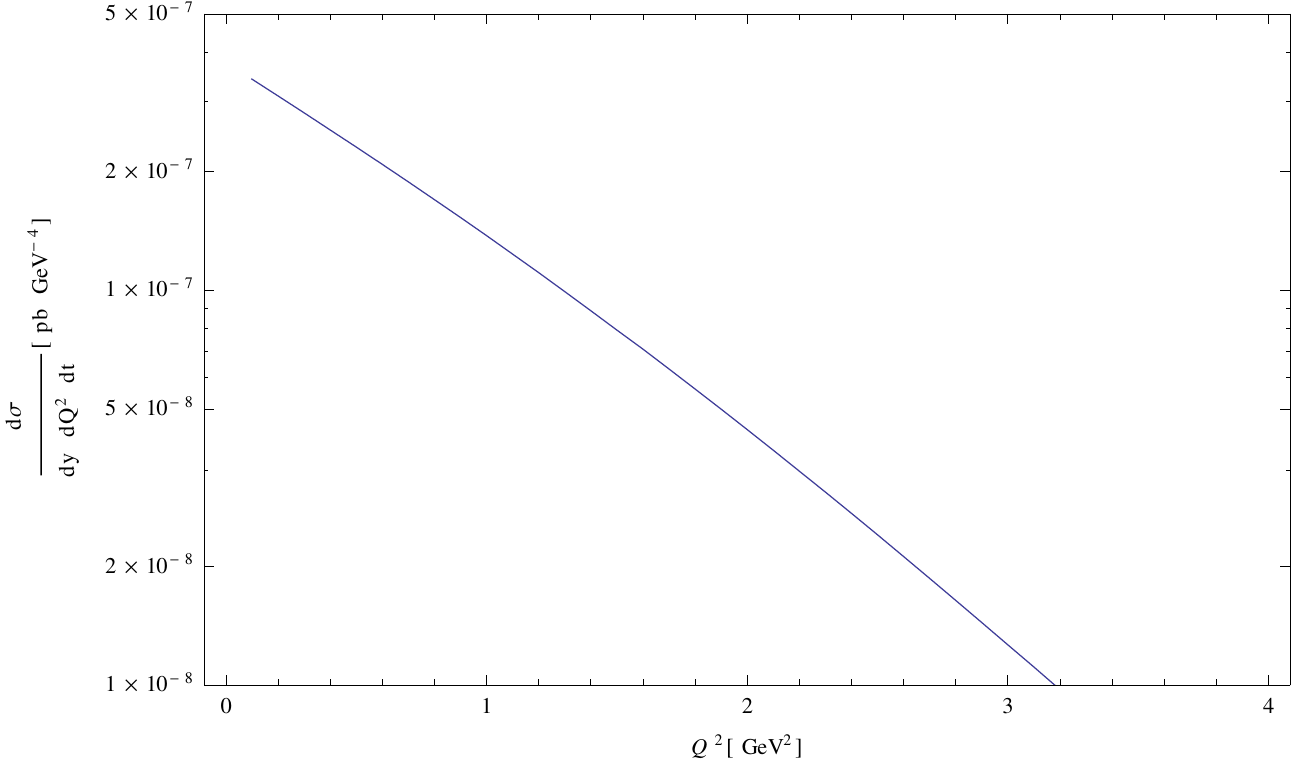}
\caption{The $Q^2$ dependence of the longitudinal contribution to the cross section $\frac{d\sigma(\bar \nu p \to l^+ n \bar D^0)}{dy\, dQ^2\, dt}$ (in pb GeV$^{-4}$) for $y=0.7, \Delta_T = 0$  and $s=20$ GeV$^2$.}
   \label{anti_sigma{00}_dy_D0}
\end{figure}

  Going from the neutrino to the antineutrino case essentially leads to a transformation $z \to \bar z$ and $x \to -x$ in the expression of the amplitude. Using the fact that $\phi^{D^-}(\bar z) = \phi^{D^+}( z)$, the amplitudes can be written in terms of the same DA, but taking the GPD as $H(-x, \xi, t), \tilde H(-x, \xi, t), ...$. For obvious reasons, the gluon contributions are the same as for the neutrino case but the quark contributions are quite different since the weak negatively charged current selects $\bar d$-antiquark rather than the $d$-quark contributions.  Moreover, there is a relative sign change between the gluon and quark contributions so that the antineutrino cross sections now read (with the same approximations as for the neutrino case):
\begin{eqnarray}
\sigma^{\bar \nu}_{00} =    \frac{1} { Q^2}\biggl\{ && [\, |C_q{\mathcal{H}}_{L} - C_g{\mathcal{H}}_{g}|^2 + |C_q\tilde{\mathcal{H}}_{L} + C_g\tilde{\mathcal{H}}_{g}|^2 ] (1-\xi^2) +\frac{\xi^4}{1-\xi^2} [\, |C_q\tilde{\mathcal{E}}_{L}+C_g\tilde{\mathcal{E}}_{g} |^2 +  |C_q {\mathcal{E}}_{L}-C_g {\mathcal{E}}_{g} |^2]  \nonumber  \\
  &&  -2 \xi^2 {\mathcal R}e  [C_q{\mathcal{H}}_{L} - C_g{\mathcal{H}}_{g}] [C_q {\mathcal{E}}^*_{L}-C_g {\mathcal{E}}^*_{g}]  -2 \xi^2 {\mathcal R}e  [C_q\tilde{\mathcal{H}}_{L} + C_g\tilde{\mathcal{H}}_{g}] [C_q \tilde{\mathcal{E}}^*_{L}+C_g \tilde{\mathcal{E}}^*_{g}] \biggr\} ,\, 
\end{eqnarray}
while $\sigma^{\bar \nu}_{--}$ is identical to the neutrino case:
\begin{eqnarray}
\sigma^{\bar \nu}_{--} =   \frac{16\xi^2 C_q^2 (m_c-2M_D)^2}{(Q^2+M_D^2)^2}\biggl\{(1-\xi^2)|{\mathcal{H}_T}|^2  + \frac{\xi^2}{1-\xi^2} | { {\mathcal{E'}}_T}|^2 -2\xi \mathcal{R}e [ \mathcal{H}_T { {\mathcal{E'}}_T^{ *}}]\biggr\} \,, 
\end{eqnarray}
 but with the chiral-odd GPDs taken as $H_T(-x, \xi, t), E'_T(-x, \xi, t), ...$ when calculating $\mathcal{H}_T , \mathcal{E'}_T ...$ 
We show on Fig.\ref{anti_sigma{--}_dy_D-_on_proton_and_neutron} the  transverse  cross sections for the production of a $D^-$ which is an order of magnitude larger for the neutron (solid curve and upper band) than for the proton target. On Fig. \ref{anti_sigma{00}_dy_D-_on_proton_neutron_quarks_gluons} the  plot of the  longitudinal cross sections for the production of a $D^-$ on a proton and on a neutron shows an important partial cancellation of the quark contribution (dashed curve) by the gluon contribution into the total (solid line) cross section. On Fig. \ref{anti_sigma{--}_D0} and Fig. \ref{anti_sigma{00}_dy_D0}, we plot  the transverse and longitudinal  cross sections for the reaction $\bar \nu p \to l^+ \bar D^0 n$ where there is no gluon contribution. The transverse cross section dominates for $Q^2 > 2$ GeV$^2$ making this process the most sensitive probe of the chiral-odd GPDs.

\section{Conclusion.}
 Collinear QCD factorization has allowed us  to calculate exclusive neutrino production of $D-$mesons in terms of GPDs. Let us stress that at medium energy, this exclusive channel should dominate $D-$meson production. Our study complements the previous calculations \cite{Kopeliovich:2012dr} which were dedicated to the production of pseudoscalar light meson, but were omitting gluon contributions.
 
We have demonstrated that gluon and both  chiral-odd and chiral-even quark GPDs contribute in specific ways to the amplitude for different polarization states of the W. The $y-$dependence of the cross section allows to separate different contributions and the measurement of the azimuthal dependence, through the moments $<cos \varphi>$ and $<sin \varphi>$ singles out the transversity chiral-odd GPDs contributions. The flavor dependence, and in particular the difference between $D^+$ and $D^0$ production rates, allows to test the importance of gluonic contributions. The behaviour of the proton and neutron target cross sections enables to separate the $u$ and $d$ quark contributions.

 Experimental data\cite{data}  already demonstrated their ability to distinguish different channels for charm production in neutrino and anti neutrino experiments. The statistics were however  too low to separate longitudinal and transverse contributions. Moreover their analysis
was not undertaken in the recent appropriate theoretical framework where skewness effects are taken into account. Planned medium and high energy neutrino facilities \cite{NOVA} and experiments such  as Miner$\nu$a \cite{Aliaga:2013uqz} and MINOS+ \cite{Timmons:2015laq} which have their scientific program oriented toward the understanding of neutrino oscillations  or to the discovery of the presently elusive sterile neutrinos will collect more statistics and will thus allow - without much additional equipment - some important progress in the realm of hadronic physics.


\paragraph*{Acknowledgements.}
\noindent
We thank Katarzyna Grzelak, Trung Le and Luis Alvarez Ruso for useful discussions and correspondence.
 This work is partly supported by grant No 2015/17/B/ST2/01838 from the National Science Center in Poland, by the Polish-French collaboration agreements  Polonium and COPIN-IN2P3, and  by the French grant ANR PARTONS (Grant No. ANR-12-MONU-0008-01).


\begin{thebibliography}{99}

  
\bibitem{fact1}
D.~M{\"u}ller {\it et al.},
Fortsch.\ Phys.\  {\bf 42},  101 (1994).

\bibitem{fact2}
X.~Ji,
  Phys.\  Rev.\ {\bf D55}, 7114 (1997);
A.~V.~Radyushkin,
Phys.\ Rev.\ D {\bf 56}, 5524 (1997).

\bibitem{fact3}
J.~C.~Collins, L.~Frankfurt, M.~Strikman,
Phys.\ Rev.\ D {\bf 56}, 2982 (1997).


\bibitem{3d}
 M.~Burkardt,
  Phys.\ Rev.\ D {\bf 62}, 071503 (2000)
  [Phys.\ Rev.\ D {\bf 66}, 119903 (2002)];
  J.~P.~Ralston and B.~Pire,
  Phys.\ Rev.\ D {\bf 66}, 111501 (2002);
 M.~Diehl and P.~Hagler,
  Eur.\ Phys.\ J.\ C {\bf 44}, 87 (2005).
  
 \bibitem{weakGPD}
  B.~Lehmann-Dronke and A.~Schafer,
  Phys.\ Lett.\ B {\bf 521} (2001) 55;
  C.~Coriano and M.~Guzzi,
  Phys.\ Rev.\ D {\bf 71} (2005) 053002;
  P.~Amore, C.~Coriano and M.~Guzzi,
  JHEP {\bf 0502} (2005) 038;
  A.~Psaker, W.~Melnitchouk and A.~V.~Radyushkin,
  Phys.\ Rev.\ D {\bf 75} (2007) 054001.
  
  \bibitem{PS} 
  B.~Pire and L.~Szymanowski,
  Phys.\ Rev.\ Lett.\  {\bf 115}, 092001 (2015).
  doi:10.1103/PhysRevLett.115.092001;
   B.~Pire, L.~Szymanowski and J.~Wagner,
  EPJ Web Conf.\  {\bf 112}, 01018 (2016)
  doi:10.1051/epjconf/201611201018
  \bibitem{extraction}
 M.~Guidal,
  Eur.\ Phys.\ J.\ A {\bf 37}, 319 (2008)
  Erratum: [Eur.\ Phys.\ J.\ A {\bf 40}, 119 (2009)];
 M.~Guidal, H.~Moutarde and M.~Vanderhaeghen,
  Rept.\ Prog.\ Phys.\  {\bf 76}, 066202 (2013);
  B.~Berthou {\it et al.},
arXiv:1512.06174 [hep-ph];
K.~Kumericki and D.~Mueller,
  EPJ Web Conf.\  {\bf 112}, 01012 (2016);
K.~Kumericki, S.~Liuti and H.~Moutarde,
  Eur.\ Phys.\ J.\ A {\bf 52}, no. 6, 157 (2016).

\bibitem{data}
N.~Ushida {\it et al.} [Fermilab E531 Collaboration],
  Phys.\ Lett.\ B {\bf 206}, 375 (1988);
S.~A.~Rabinowitz {\it et al.},
  Phys.\ Rev.\ Lett.\  {\bf 70}, 134 (1993);
P.~Vilain {\it et al.} [CHARM II Collaboration],
  Eur.\ Phys.\ J.\ C {\bf 11}, 19 (1999);
P.~Astier {\it et al.} [NOMAD Collaboration],
  Phys.\ Lett.\ B {\bf 486}, 35 (2000);
 A.~Kayis-Topaksu {\it et al.},
  New J.\ Phys.\  {\bf 13}, 093002 (2011)
  [arXiv:1107.0613 [hep-ex]].

  
  \bibitem{MD}
  M.~Diehl,
  Phys.\ Rept.\  {\bf 388} (2003) 41.
  
  \bibitem{DDVCS} 
  M.~Guidal and M.~Vanderhaeghen,
  Phys.\ Rev.\ Lett.\  {\bf 90}, 012001 (2003).
  doi:10.1103/PhysRevLett.90.012001


\bibitem{TCS}
  E.~R.~Berger, M.~Diehl and B.~Pire,
  Eur.\ Phys.\ J.\ C {\bf 23} (2002) 675.
  doi:10.1007/s100520200917
  
      \bibitem{heavyDA} 
  A.~Szczepaniak, E.~M.~Henley and S.~J.~Brodsky,
  Phys.\ Lett.\ B {\bf 243}, 287 (1990);

   S.~Descotes-Genon and C.~T.~Sachrajda,
  Nucl.\ Phys.\ B {\bf 650}, 356 (2003);
  V.~M.~Braun, D.~Y.~Ivanov and G.~P.~Korchemsky,
  Phys.\ Rev.\ D {\bf 69}, 034014 (2004);
  T.~Feldmann, B.~O.~Lange and Y.~M.~Wang,
  Phys.\ Rev.\ D {\bf 89}, no. 11, 114001 (2014);
   V.~M.~Braun and A.~Khodjamirian,
  Phys.\ Lett.\ B {\bf 718}, 1014 (2013).

 \bibitem{heavyDA2} 
   T.~Kurimoto, H.~n.~Li and A.~I.~Sanda,
  Phys.\ Rev.\ D {\bf 65}, 014007 (2002);
  T.~Kurimoto, H.~n.~Li and A.~I.~Sanda,
  Phys.\ Rev.\ D {\bf 67}, 054028 (2003).
  doi:10.1103/PhysRevD.67.054028


 \bibitem{transGPDdef}
 M.~Diehl,
  Eur.\ Phys.\ J.\  C {\bf 19}, 485 (2001).
  
  \bibitem{transGPDno}
  M.~Diehl {\em et. al.}
  Phys.\ Rev.\  D {\bf 59}, 034023 (1999);
  J.~C.~Collins {\em et. al.},
  Phys.\ Rev.\  D {\bf 61}, 114015 (2000).

\bibitem{transGPDacc}
 D.~Yu.~Ivanov  {\it et al.},
  Phys.\ Lett.\  B {\bf 550}, 65 (2002);
R.~Enberg, B.~Pire and L.~Szymanowski,
  Eur.\ Phys.\ J.\  C {\bf 47}, 87 (2006);
 M.~E.~Beiyad  {\it et al.},
 Phys.\ Lett.\ B {\bf 688}, 154 (2010);
  R.~Boussarie, B.~Pire, L.~Szymanowski and S.~Wallon,
  JHEP {\bf 1702}, 054 (2017)
  doi:10.1007/JHEP02(2017)054
 
  
  \bibitem{CEGPD}
  S.~V.~Goloskokov and P.~Kroll,
  Eur.\ Phys.\ J.\ C {\bf 42}, 281 (2005);
  S.~V.~Goloskokov and P.~Kroll,
  Eur.\ Phys.\ J.\ C {\bf 50}, 829 (2007);
   S.~V.~Goloskokov and P.~Kroll,
  Eur.\ Phys.\ J.\ C {\bf 53}, 367 (2008);
   P.~Kroll, H.~Moutarde and F.~Sabatie,
  Eur.\ Phys.\ J.\ C {\bf 73}, no. 1, 2278 (2013).
  
  \bibitem{Goloskokov:2011rd}
  S.~V.~Goloskokov and P.~Kroll,
  Eur.\ Phys.\ J.\ A {\bf 47} (2011) 112
  [arXiv:1106.4897 [hep-ph]].
 
  \bibitem{models}
   B.~Pire, K.~Semenov-Tian-Shansky, L.~Szymanowski and S.~Wallon,
  Eur.\ Phys.\ J.\ A {\bf 50} (2014) 90;
  X.~Xiong and J.~H.~Zhang,
  Phys.\ Rev.\ D {\bf 92}, no. 5, 054037 (2015).
  
  \bibitem{lattice}
M.~Gockeler {\it et al.} [QCDSF and UKQCD Collaborations],
  Phys.\ Rev.\ Lett.\  {\bf 98}, 222001 (2007).

   \bibitem{Collins:1998rz} 
  J.~C.~Collins,
  Phys.\ Rev.\ D {\bf 58}, 094002 (1998).
    
    
\bibitem{Arens} 
 see for instance  T.~Arens, O.~Nachtmann, M.~Diehl and P.~V.~Landshoff,
  Z.\ Phys.\ C {\bf 74}, 651 (1997).

  \bibitem{Kopeliovich:2012dr} 
  B.~Z.~Kopeliovich, I.~Schmidt and M.~Siddikov,
  Phys.\ Rev.\ D {\bf 86}, 113018 (2012) and
 D {\bf 89}, 053001 (2014);
   G.~R.~Goldstein, O.~G.~Hernandez, S.~Liuti and T.~McAskill,
  AIP Conf.\ Proc.\  {\bf 1222} (2010) 248;
   M.~Siddikov and I.~Schmidt,
  arXiv:1611.07294 [hep-ph].

  
  \bibitem{DGPR}
  M.~Diehl  {\it et al.},
  Phys.\ Lett.\ B {\bf 411} (1997) 193;
   A.~V.~Belitsky, D.~Mueller and A.~Kirchner,
  Nucl.\ Phys.\ B {\bf 629}, 323 (2002).

 \bibitem{Ren:2017xov} 
  L.~Ren {\it et al.} [Minerva Collaboration],
  arXiv:1701.04857 [hep-ex].
    
  \bibitem{NOVA} 
  D.~S.~Ayres {\it et al.}  [NOvA Collaboration],
  hep-ex/0503053; see also
   M.~L.~Mangano, S.~I.~Alekhin, M.~Anselmino {\it et al.},
  CERN Yellow Report CERN-2004-002, pp.185-257
  [hep-ph/0105155].


\bibitem{Aliaga:2013uqz} 
  L.~Aliaga {\it et al.} [MINERvA Collaboration],
  Nucl.\ Instrum.\ Meth.\ A {\bf 743}, 130 (2014)
  [arXiv:1305.5199 [physics.ins-det]]; 
   J.~Devan {\it et al.} [MINERvA Collaboration],
  Phys.\ Rev.\ D {\bf 94}, 112007 (2016),
  [arXiv:1610.04746 [hep-ex]]; 
   C.~L.~McGivern {\it et al.} [MINERvA Collaboration],
  Phys.\ Rev.\ D {\bf 94}, no. 5, 052005 (2016),
  [arXiv:1606.07127 [hep-ex]]; 
   C.~M.~Marshall {\it et al.} [MINERvA Collaboration],
  [arXiv:1611.02224 [hep-ex]].
 
\bibitem{Timmons:2015laq} 
  A.~Timmons,
  Adv.\ High Energy Phys.\  {\bf 2016}, 7064960 (2016)
  [arXiv:1511.06178 [hep-ex]].
 
  
\end{thebibliography}
\end{document}